\documentclass{aa}

\usepackage[hidelinks]{hyperref}
\usepackage[varg]{txfonts}
\usepackage{natbib}
\usepackage{graphicx}
\usepackage{gensymb}
\usepackage{textcomp}
\usepackage{pdflscape}
\bibpunct{(}{)}{;}{a}{}{,} % to follow the A&A style

\begin{document}
\title{Studying bright variable stars with\\ the Multi-site All-Sky CAmeRA (MASCARA)\thanks{Tables~\ref{t:p} and~\ref{t:n} are also available in electronic form at the CDS via anonymous ftp to cdsarc.u-strasbg.fr (130.79.128.5) or via \url{http://cdsweb.u-strasbg.fr/cgi-bin/qcat?J/A+A/}.}}

\author{O. Burggraaff \inst{1} \thanks{\email{burggraaff@strw.leidenuniv.nl}}
   \and G.J.J. Talens \inst{1}
   \and J. Spronck    \inst{1}
   \and A.-L. Lesage  \inst{1}
   \and R. Stuik      \inst{1}
   \and G.P.P.L. Otten\inst{1}
   \and V. Van Eylen  \inst{1}
   \and D. Pollacco   \inst{2}
   \and I.A.G. Snellen\inst{1}}

\institute{Leiden Observatory, Leiden University, PO Box 9513, 2300 RA Leiden, The Netherlands \and Department of Physics, University of Warwick, Coventry CV4 7AL, UK}

\date{Received day month year / Accepted day month year}

\abstract
{The Multi-site All-Sky CAmeRA (MASCARA) aims to find the brightest transiting planet systems by monitoring the full sky at magnitudes $4 < V < 8.4$, taking data every 6.4 seconds. The northern station has been operational on La Palma since February 2015. These data can also be used for other scientific purposes, such as the study of variable stars.}
{In this paper we aim to assess the value of MASCARA data for studying variable stars by determining to what extent known variable stars can be recovered and characterised, and how well new, unknown variables can be discovered.}
{We used the first 14 months of MASCARA data, consisting of the light curves of 53\,401 stars with up to one million flux points per object. All stars were cross-matched with the VSX catalogue to identify known variables. The MASCARA light curves were searched for periodic flux variability using generalised Lomb-Scargle periodograms. If significant variability of a known variable was detected, the found period and amplitude were compared with those listed in the VSX database. If no previous record of variability was found, the data were phase folded to attempt a classification.}
{Of the 1919 known variable stars in the MASCARA sample with periods $0.1 < P < 10$ days, amplitudes $> 2\%$, and that have more than $80$ hours of data, $93.5\%$ are recovered. In addition, the periods of 210 stars without a previous VSX record were determined, and 282 candidate variable stars were newly identified. We also investigated whether second order variability effects could be identified. The O’Connell effect is seen in seven eclipsing binaries, of which two have no previous record of this effect.}
{MASCARA data are very well suited to study known variable stars. They also serve as a powerful means to find new variables among the brightest stars in the sky. Follow-up is required to ensure that the observed variability does not originate from faint background objects.}


\maketitle

\section{Introduction} \label{s:intro}

Variable stars--stars that change in magnitude over time--have been a field of study since antiquity \citep{algolegypt}. Currently, over 500\,000 examples are listed in the International Variable Star Index\footnote{\url{https://www.aavso.org/vsx/}} (VSX). Variable stars are often discovered as a secondary science goal of large stellar survey projects, such as OGLE \citep{OGLE2, OGLE_ceps} and the NASA Kepler Mission \citep{Kepler_EBs}, which each have identified thousands of variable stars albeit at relatively faint magnitudes. Astrometric surveys such as Hipparcos \citep{HIP_vars} and currently Gaia \citep{Gaia_DR1} perform all-sky surveys that include the brightest stars, but only with a relatively low number of measurements per object. A number of other surveys have identified bright variable stars, such as ASAS \citep{ASAS}, KELT \citep{KELT} and MOST \citep{MOST}. TESS will provide excellent photometry on stars as bright as $V = 4.5$ mag but will be limited by its mission duration to relatively short period ($P_\text{max} \approx 40$ days) variable stars only \citep{TESS}. Additionally, the study of variable stars is one of the richest fields in terms of amateur contributions. Organisations such as the American Association of Variable Star Observers (AAVSO) provide light curves of thousands of stars over periods of decades, based in part on volunteer work. However, coverage is mostly sparse and heterogeneous.

Variable stars have great value across many fields of astrophysics. Pulsating variable stars such as cepheids, have been used to accurately determine distances of deep-sky objects \citep{hubble}. Currently these stars, and other types of variables in the `instability strip' on the Hertzsprung-Russell diagram, are often used as testing grounds for models of stellar structure and evolution \citep{instability1, instability2, instability3}. Eclipsing binary systems provide measurements of the masses and radii of their components to the level of accuracy needed to constrain models of stellar structure. Since any type of star can be part of a binary system, this method allows for measurements of these parameters across the Hertzsprung-Russell diagram, rather than only specific sections of it \citep{EBreview}. Space missions such as BRITE are now capable of observing many such stars with high precision, short cadence and over long time scales \citep{brite}.

In this paper we wish to assess how valuable Multi-site All-Sky CAmeRA (MASCARA) data are to study variability in bright stars. As far as we know, MASCARA is currently the only survey that monitors the near-entire sky at $V < 8$ magnitudes. In Sect.~\ref{s:data} we discuss MASCARA and its data. In Sect.~\ref{s:meth} the analysis is presented, in which we determine the recovery rate of known variable stars in the first 14 months of data. Furthermore, the MASCARA data is searched for new yet-unknown variables. In Sect.~\ref{s:res} the results are presented, which are discussed in Sect.~\ref{s:dis}.

\section{MASCARA data} \label{s:data}

The main goal of the Multi-Site All-Sky CAmeRA (MASCARA) is to detect exoplanets around bright stars using the transit method. The northern-hemisphere MASCARA station, located on La Palma (Canary Islands, Spain), has been fully operational since February 2015. The southern station, located at La Silla (Chile), saw its first light in June 2017. Thus far, two exoplanets have been discovered using MASCARA data \citep{MASCARA1b, MASCARA2b}.

Each MASCARA station contains five cameras, one pointed in each cardinal direction and one at zenith, covering the local sky down to airmass two to three. The cameras are modified Atik 11000M interline CCDs, without a filter, giving them a spectral range of approximately 300 to 1000 nanometres. Each camera is equipped with a Canon 24 mm $f/1.4$ USM L II lens with a 17 mm aperture, providing a $53\degree \times 74\degree$ field of view each, at a scale of approximately 1 arcminute per pixel. For a detailed description we refer the reader to \citet{MASCARA2017}.

The cameras take 6.4 second back-to-back exposures through the night at fixed local sidereal times. Aperture photometry is applied to these images to extract the fluxes of all the stars with $V < 8.4$ mag. This is done automatically for a list of stars known to be visible with MASCARA, based on the All-Sky Compiled Catalogue \citep[ASCC;][]{ascc}. These measurements are binned in groups of 50, producing a light curve with a binned data point every 320 seconds. A detailed description of the MASCARA data reduction pipeline and analysis is presented in Talens et al. (in prep.).

Our analysis is based on the first year of data of the northern station, taken between February 2015 and March 2016 (heliocentric Julian dates (HJD) 2\,457\,056 -- 2\,457\,480). The data set consists of up to 25\,000 binned data points (HJD, magnitude, magnitude error) per star, with a median number of binned data points of 12\,757 ($\approx 1100$ hours). The number of flux points for a given star depends mainly on its sky coordinates, in particular its declination. The stars range in right ascension from $0^h$ to $24^h$, in declination from $-38.6\degree$ to $+90\degree$, and in $V$-magnitude from $2.0$ to $8.4$. We note that the brightest stars, with $V < 4$ mag, are likely to be saturated at certain parts of the CCDs, effectively reducing the number of usable data.

\section{Analysis} \label{s:meth}

Our MASCARA data analysis consists broadly of four steps, applied to each star individually. Firstly, an ansatz period of variability was searched for through the generalised Lomb-Scargle periodogram \citep[GLS;][]{GLS} of the MASCARA light curves. Secondly, systematic effects caused by the instrument and the Moon were removed. Thirdly, a direct $\chi^2$ minimalisation was performed to obtain the final estimate for the period of variability. Finally, the candidate variable star was checked for being a false positive caused by variability of a known background star. All analysis was performed using designated python scripts.

\subsection{Step 1: Finding the Ansatz period} \label{ss:gls}

\begin{figure*}
	\centering
	\includegraphics[width=17cm]{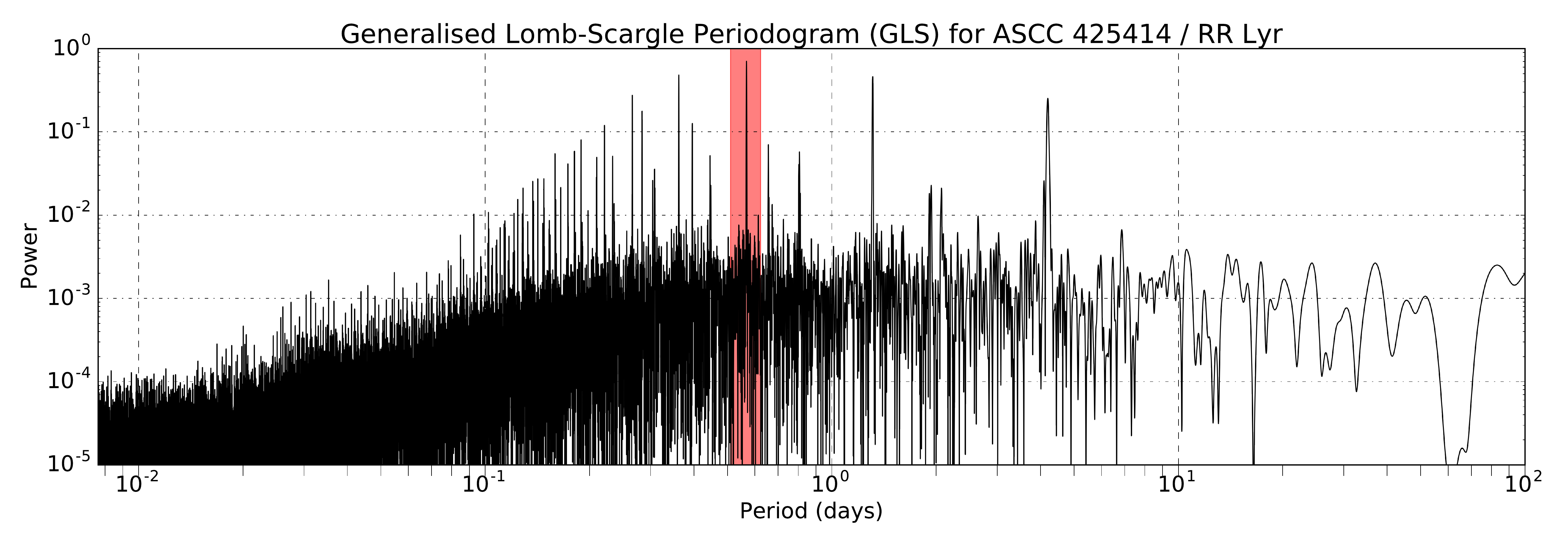}
	\caption{Generalised Lomb-Scargle periodogram of ASCC 425414 (RR Lyrae). The strongest signal is at a period of $P = 0.567$ days, highlighted with a red background. We note that in this particular case, the instrumental effects discussed in Sect.~\ref{ss:detrend} are not strongly present in the GLS.}
	\label{f:lsp}
\end{figure*}

A first estimate for the strongest periodic signal in a light curve was determined through the generalised Lomb-Scargle periodogram \citep[GLS;][]{GLS}, a variation on the standard Lomb-Scargle periodogram \citep{scargle} which allows weighting of data points and fitting of the mean value. This periodogram is equivalent to a $\chi^2$ fit of sine waves to the data. We tested up to 68\,000 periods ranging from 640 seconds to 100 days. The upper limit was set to ensure the presence of multiple cycles in the data. GLS power can range between 0 and 1, equivalent to no fit and a perfect sinusoidal fit respectively. The strongest signal in the GLS was used as a first estimate for the true period of variability. Care was taken to ignore signals within 5\% of 1 sidereal day or an alias thereof ($1/2$, $1/3$, ... days) and within 5\% of 29.5 days, as these are caused by systematic effects as described in step 2.

As an example, the GLS of ASCC 425414 (RR Lyrae) is shown in Fig.~\ref{f:lsp}. The forest of strong signals are all aliases ($f_\text{alias} = f_0 + k$ with $f_0 = 1/P_0$ and $k$ an integer) and harmonics ($P_\text{harm} = k P_0$ or $P_\text{harm} = P_0 / k$ with $k$ an integer) of the true period of $P_0 = 0.567$ days, which itself has the strongest power in the GLS diagram of $p_\text{max} \approx 0.7$.

\subsection{Step 2: Removal of systematics} \label{ss:detrend}

Two important systematic effects are present in the MASCARA light curves. The first has a period of one sidereal day and is caused by the varying PSF of the cameras across their field of view. The second has a period of 29.5 days and is caused by changing background levels due to the Moon. The amplitude and significance of these effects differ between stars and are related to their sky position, magnitude and the amplitude of their variability.

\begin{figure}
	\centering
	\includegraphics[width=\columnwidth]{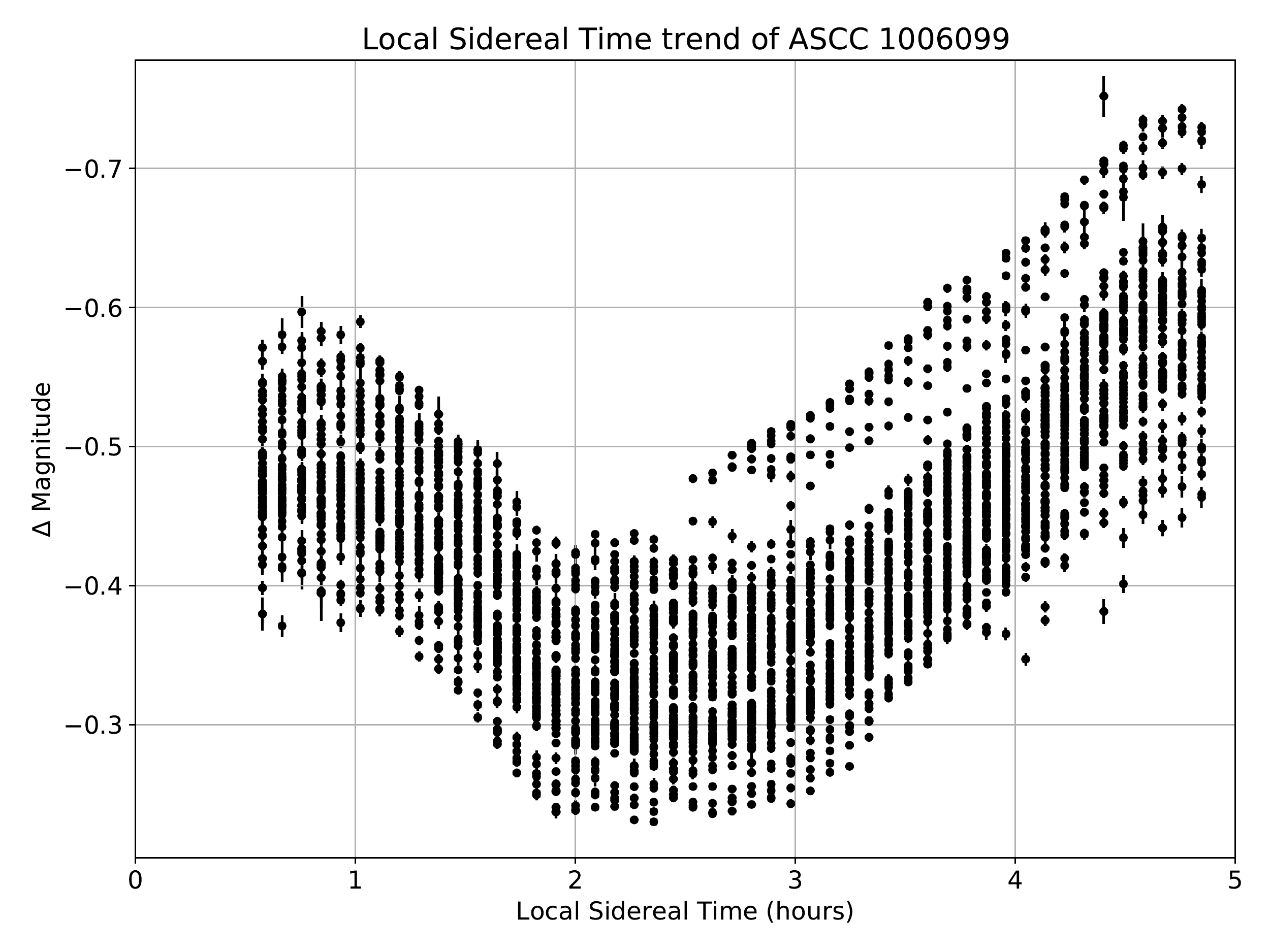}
	\caption{PSF variations in the zenith camera data for the star ASCC 1006099. The magnitude axis zero-point is arbitrary. We note that the data are taken at fixed sidereal times.}
	\label{f:lst}
\end{figure}

Systematic flux variations with a period of 1 sidereal day are caused by the considerably variable point spread function of a MASCARA camera across its field of view \citep[see][]{MASCARA2017}. Since the cameras stare at a fixed position, a star typically travels across the CCD in a few hours, significantly changing the fraction of light that falls within the aperture used to obtain the photometric measurements. Since all flux measurements are obtained relatively to a set of surrounding stars, to first order this effect cancels out. However, since the PSF changes so strongly, faint wings from neighbouring stars enter and leave the aperture according to the position of the star on the CCD -- an effect that is unique to each individual object. It will be the strongest for faint stars with very close and bright neighbours and can have amplitudes up to 0.5 in magnitude. Fortunately, the path of a star on the CCD is nearly identical for every sidereal day, and therefore this systematic effect can be measured and removed. An example of such an LST (local sidereal time) trend for the star ASCC 1006099 (EG Cet) is given in Fig.~\ref{f:lst}.

The 29.5-day effect is caused by the Moon. As the Moon moves across the sky, it significantly affects the local sky. In a way that is not yet completely understood by the MASCARA team, the sky background level influences the measured fluxes, depending on the magnitude of the star. It typically has an amplitude of 0.01 magnitude. Additionally ghost images will appear. These effects are difficult to predict and thus are not removed in the original data pipeline. The resulting effect also has a period of 29.5 days and can have significant amplitudes of up to 0.6 magnitude.

To remove these systematics, firstly the data were phase folded with the ansatz period determined from the GLS (Step 1). The resulting light curve was binned in phase space using 150 bins, and the weighted mean of each bin was removed from the data. The residuals subsequently contain{ed} only the LST and lunar trends. First the residuals were phase folded with $P = 29.5$ days, again binned in phase space, and that resulting trend was removed from the original data. The process was then repeated for $P = 1$ sidereal day, removing that trend from the data as well. This was done iteratively until the LST and lunar trends were below 0.001 magnitude in amplitude. The systematics were removed on a per-camera basis while the ansatz period phase-fold was done with data from all cameras combined. The resulting detrended data were used for further analysis.

\subsection{Step 3: Final period estimate} \label{ss:chi2}

In the next step the GLS was calculated again for the detrended data, and its strongest period determined. This was generally very close to the ansatz period from the original GLS, but for a small number of stars the period with the strongest signal after detrending changed -- now in line with the literature value. Also, since in general the light curves do not resemble sinusoids, the period determined from the GLS may differ slightly from the real period.

We therefore repeat{ed} the phase fold and binning procedure from Step 2 with the detrended data for 1000 periods in the $\pm$0.5\% range around the GLS period $P_{GLS}$, in addition to similar ranges around  $2 P_{GLS}$ and $4 P_{GLS}$ to search possibly better solutions at twice and four times the period. This is important mostly for eclipsing binaries, for which the light curve is more similar to a sine wave (and thus appears stronger in the GLS) when the primary and secondary eclipse are overlaid on each other. In general, only $2 P_{GLS}$ and $4 P_{GLS}$ were tested because in a sub-sample, no stars were found to have a stronger signal for $0.5 P_{GLS}$ or other multiples. Since this $\chi^2$ calculation is a computationally expensive operation, it was chosen to only do $1$, $2$ and $4 P_{GLS}$. The final period was chosen to be that with the lowest $\chi^2$ of the phase-folded binned data points with respect to the binned-averaged light curve. The uncertainty interval on the final period estimate was determined from the $\chi^2$ curve using standard methods. 

For a small sub-set of the new variable star candidates and known variable stars with new parameters determined by MASCARA, namely 26 out of 492 stars, manual adjustment of the period was necessary. These were generally long-period variables of which the period needed to be halved and eclipsing binaries with elliptical orbits, which cause a phase difference between primary and secondary eclipse significantly less than 0.5. For these stars, the range for the $\chi^2$ calculation was manually adjusted based on a visual inspection of the light curve.

\subsection{Step 4: Removal of false positives} \label{ss:false_pos}

\begin{figure}
	\centering
	\includegraphics[width=\columnwidth]{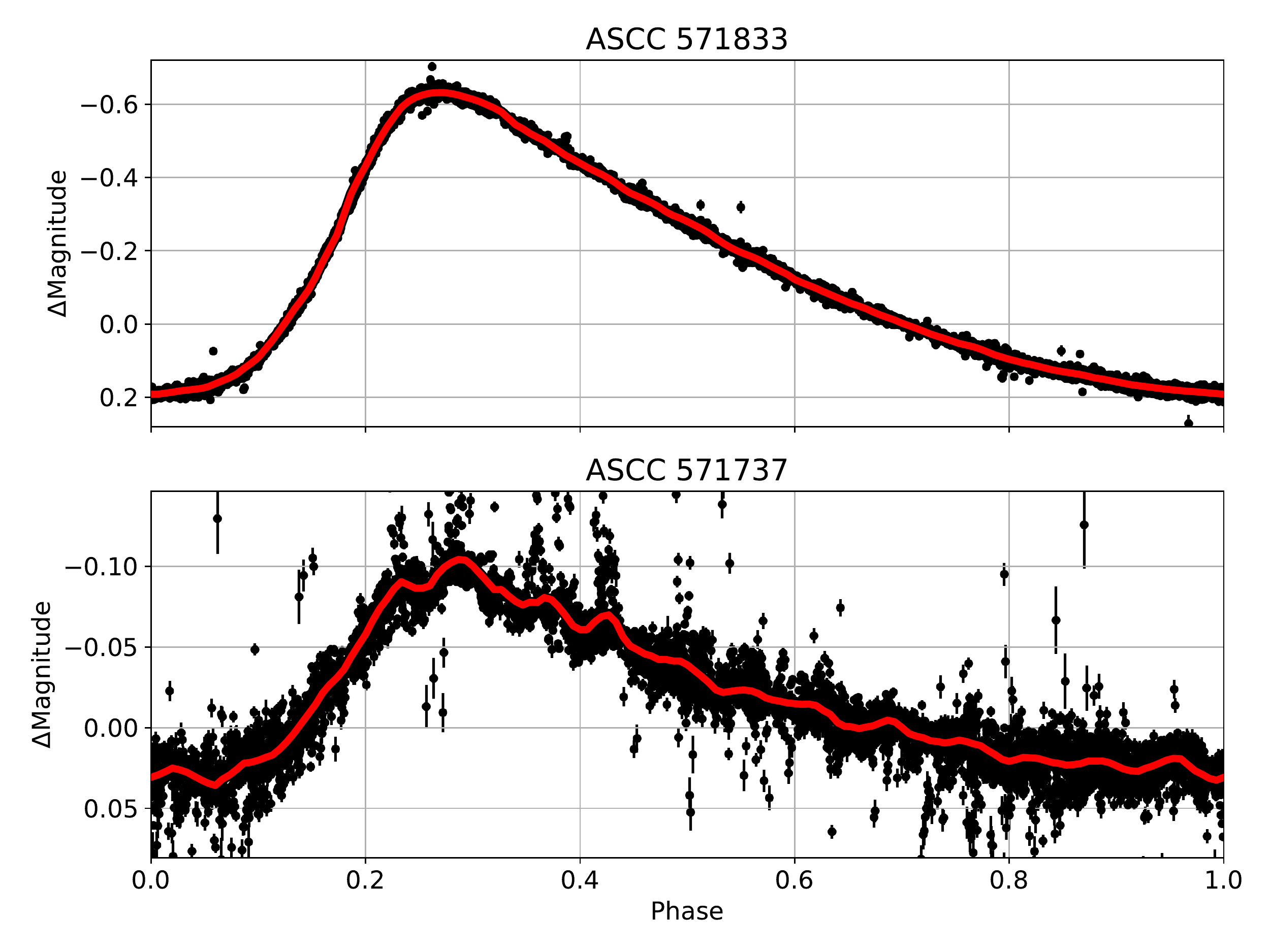}
	\caption{Comparison between light curves of ASCC 571833 (top; RT Aur) and ASCC 571737 (bottom; HD 45237), both phase folded with the same period $P = 3.72816$ days. The red line is the weighted mean; the magnitude axis zero-points are arbitrary. The period and shape of the light curves are similar but the amplitudes are different: 0.82 vs. 0.13 magnitude. The variability seen in ASCC 571737 is completely caused by that in ASCC 571833.}
	\label{f:fpos}
\end{figure}

Due to the low resolution of the MASCARA cameras (1 arcminute per pixel), there is a large degree of blending. This involves the PSFs of two stars overlapping, causing variability from one star to appear in the light curve of the other. For example, ASCC 571737, which is located 14.7\arcmin{} from ASCC 571833 (\object{RT Aur}), has in its light curve an oscillation very similar to that of RT Aur. Both light curves are shown in Fig.~\ref{f:fpos} for comparison. This blending can lead to false positive detections.

To potentially mitigate this, all known variable stars within a 1\degree{} radius of a candidate were examined. If any had a period that is similar to the candidate and a magnitude $m < 12$ in the VSX catalogue, the candidate was rejected. The magnitude limit was chosen to prevent extremely faint variables from causing false negatives. The limit is significantly lower than the lowest magnitude stars MASCARA can detect, but accounts for the heterogeneous nature of the VSX catalogue, which lists magnitudes in various bands.

\section{Results} \label{s:res}

The analysis was first tested on a sample of 2776 known variable stars with recorded periods and amplitudes in the VSX database, after which it was applied to the remainder of the stars in the MASCARA sample. The cross matching between VSX and ASCC was done by finding stars with coordinates within 10\arcsec{} from each other. 

\subsection{Recovery of known variable stars} \label{ss:acc}

\begin{figure*}
	\centering
	\includegraphics[width=17cm]{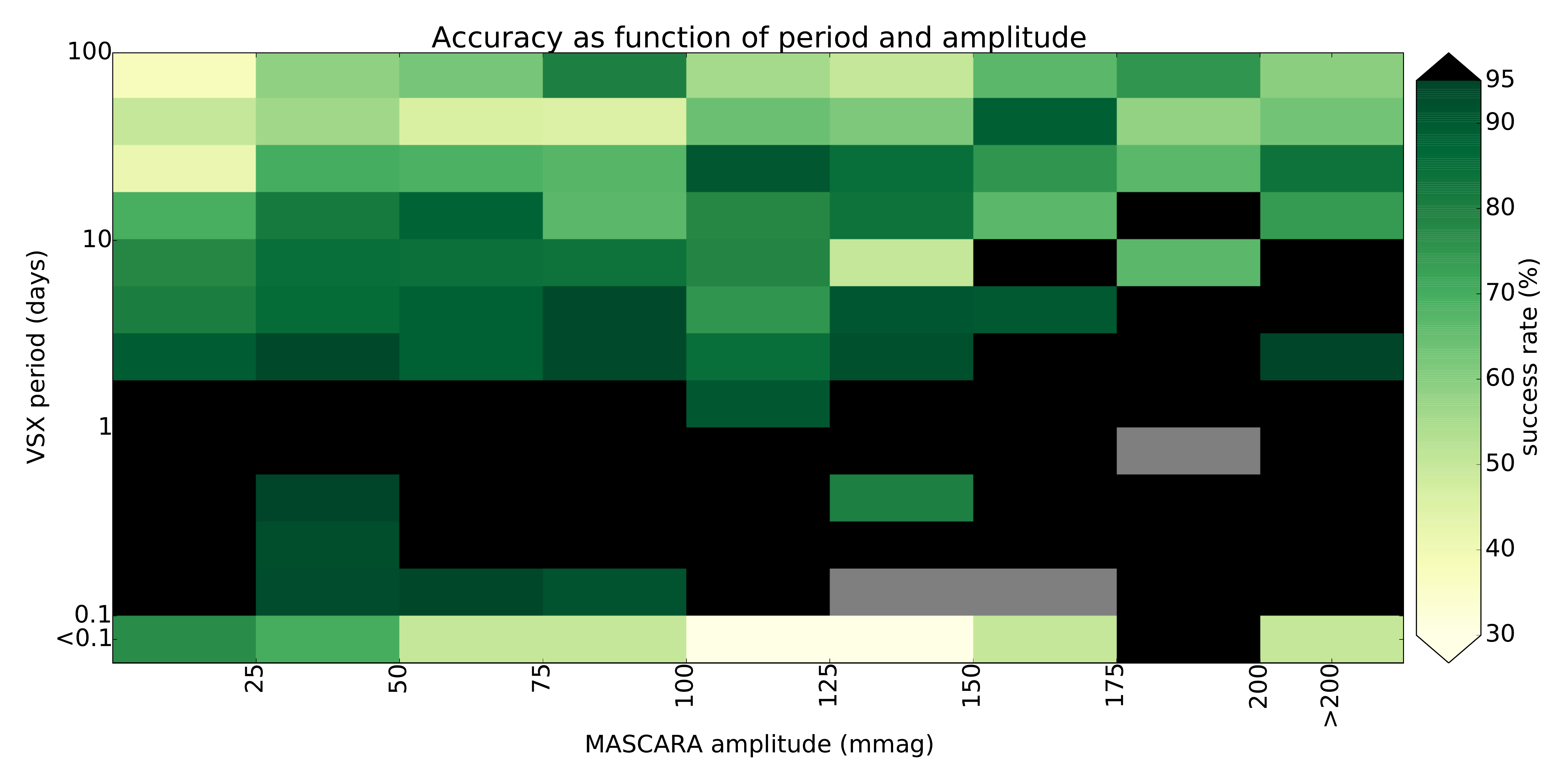}
	\caption{{R}ecovery rate of known variable stars in the MASCARA data as a function of period and amplitude of the variability. Black rectangles have a success rate $>95 \%$; grey rectangles contain no stars.}
	\label{f:suc}
\end{figure*}

The recovery rate of known variable stars from the VSX database in the MASCARA data is shown in Fig.~\ref{f:suc}. It depends strongly on the variability amplitude and period. For those stars with periods $0.1 < P < 10$ days and amplitudes $> 2$\% and $> 1000$ binned data points, 93.5\% of the objects are recovered in the first year of MASCARA data. For 40\% of the recovered objects, the catalogue and MASCARA periods match within 5\%. For amplitudes between 1 and 2\%, MASCARA finds 86.2\% of the known variables. Of the long period variable stars with $10 < P < 100$ days, MASCARA recovers 68.3\% of those known in the VSX catalogue.

\begin{figure}
	\centering
	\includegraphics[width=\columnwidth]{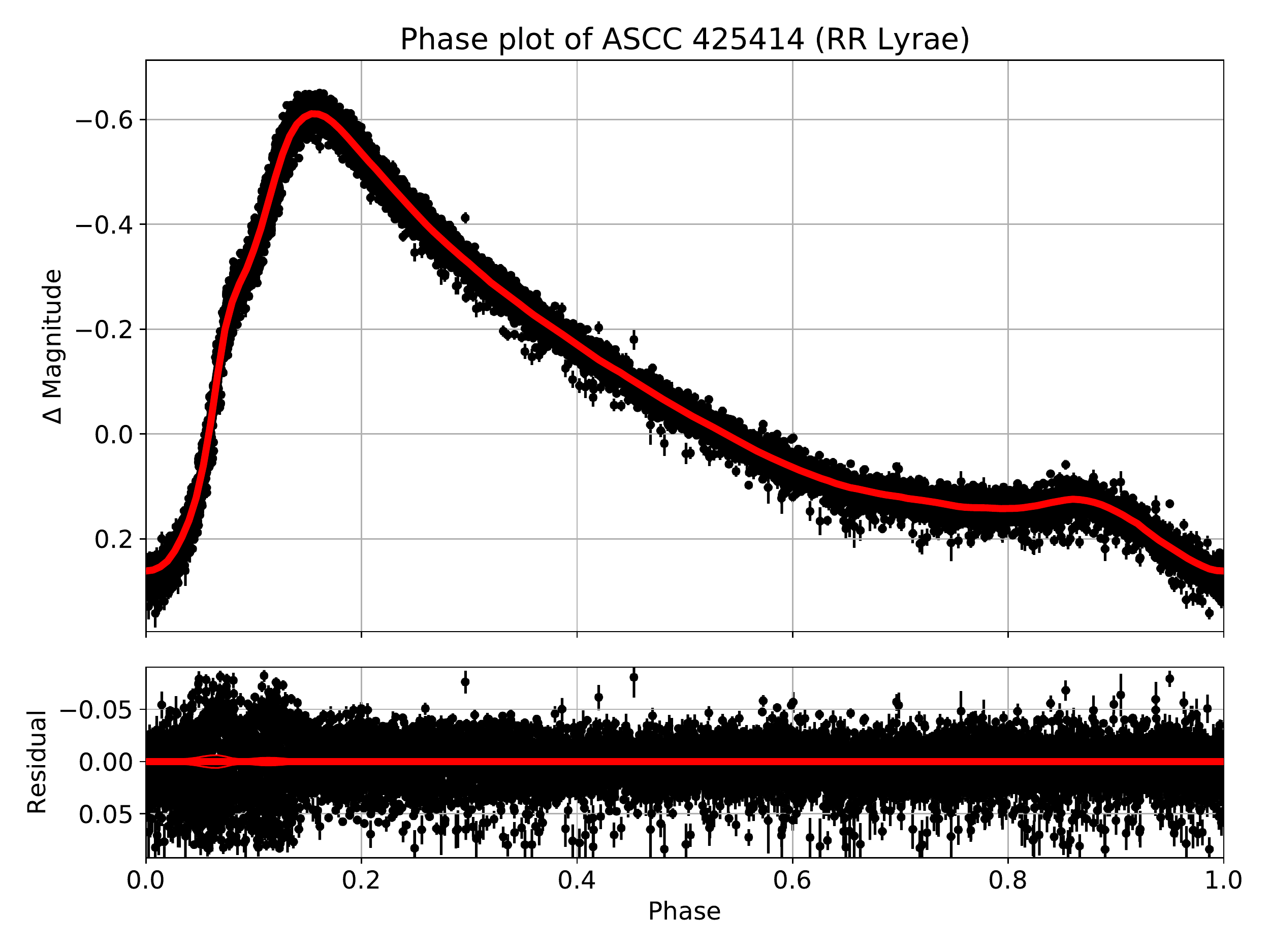}
	\caption{MASCARA light curve of 13\,279 binned data points of RR Lyrae phase folded with a period of $0.566774$ days. The red line is the running average over an 0.025 phase interval. For clarity, the data are clipped at 3-$\sigma$ from the running mean, removing 2.1\% of the binned data points.}  
	\label{f:rr}
\end{figure}

The median uncertainty in the final MASCARA period as determined in (Step 3) is three minutes, with a median relative uncertainty of 0.1\%. The found uncertainties in the period can be as low as 0.5 seconds for regular high-amplitude variable stars. The phase folded light curve of \object{RR Lyr} is given in Fig.~\ref{f:rr} as an example of the quality of MASCARA data and the period fitting. The distribution of the residuals is best fit with a Gaussian with $\sigma = 0.028$ magnitude, indicative of the typical uncertainty in the MASCARA fluxes for this star.

\subsection{New parameters for known and suspected variables} \label{ss:param}

A further 4236 stars listed in the VSX without a recorded period were analysed. Reliable periods were found for 210 of these, which are listed in Table~\ref{t:p}, with the parameters of the star (identification, coordinates, V-magnitude, number of observations by MASCARA) and of its variability (period, amplitude, epoch, VSX variability type designation). For a subset of these stars, an estimate from the MASCARA light curve of the type of variability  eclipsing binary, pulsating, or other) is also included. Light curves and periodograms of seven example stars are shown in Appendix~\ref{a:p}, and can be found for all stars with new parameters at \url{https://home.strw.leidenuniv.nl/~burggraaff/MASCARA_variables/}.

One interesting example of a previously suspected variable star recovered with MASCARA is ASCC 408281 (\object{HD 101207}). HD 101207 is a known binary system, consisting of a component A with $V_A = 8.11$ mag and a component B with $V_B = 9.32$ mag, with a visual separation of 1.97 arcsec \citep{tychodoubles}. This separation is much smaller than the MASCARA pixel size (which is approximately 1 arcmin), so the two stars are fully blended in the MASCARA data. The system has an orbital period of approximately 4000 years \citep{HD101207orbit}. \object{HD 101207B} is identified in the ASCC-VSX cross match with the suspected variable star \object{NSV 5279}.

\begin{figure}
	\centering
	\includegraphics[width=\columnwidth]{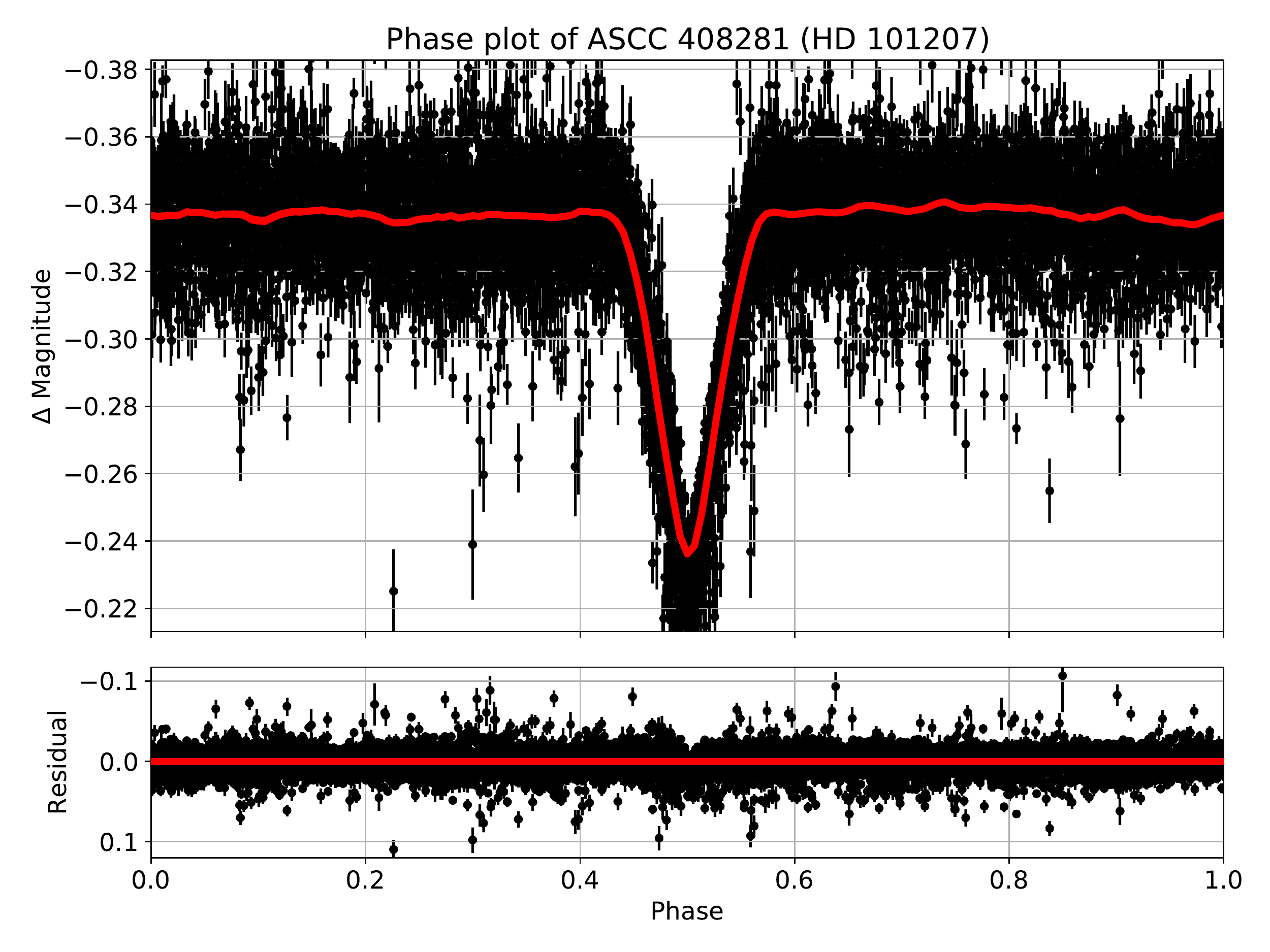}
	\caption{MASCARA light curve of ASCC 408281 (\object{HD 101207}), a previously suspected variable, which was detected with MASCARA, with a period of $1.09014(5)$ days. The red line is the running average over an 0.025 phase interval. The light curve clearly shows a single eclipse-like feature.}
	\label{f:408281}
\end{figure}

The MASCARA data for HD 101207 show a clear periodicity, with a best period estimate of $P = 1.09014(5)$ days. The phase folded light curve is given in Fig.~\ref{f:408281}. This light curve clearly shows a single dip with an amplitude of $104$ mmag, and the system can be easily identified as an eclipsing binary (EB). This feature cannot be explained by the previously known double nature of the HD 101207 system, since the orbital period of the A and B components is 4000 years.

As noted in Sect.~\ref{ss:false_pos}, blending due to the low resolution of the MASCARA cameras can cause false positive detections. However, only two systems with $V < 10$ mag were found within a 10 arcmin radius from HD 101207: BD+42 2231 and BD+41 2217. The latter has a separation from HD 101207 of 4.5 arcmin and is too faint ($V = 9.20$ mag) to cause significant blending effects at that separation. \object{BD+42 2231} has a separation from HD 101207 of 2.2 arcmin and a $V$-magnitude of $9.10$. The small separation (2 pixels) means some blending between the stars will occur, but due to the faintness of BD+42 2231, it is very unlikely to induce a variability with an amplitude as high as $104$ mmag in HD 101207. BD+42 2231 is a known spectroscopic binary, but its orbital period is $951.5 \pm 2.1$ days \citep{SBorbit}, too long to explain the variability seen in HD 101207.

Thus the most likely explanation for the observed variability is a previously unknown binary nature of one of the components of HD 101207. Since component A is significantly brighter than B ($V_A = 8.11$ mag compared to $V_B = 9.32$ mag), it is likely that \object{HD 101207A} is the eclipsing binary component. However, this cannot be said with certainty due to the complete blending of the two components in the MASCARA data. We expect that the MASCARA data contain many of such as yet undiscovered systems.

\subsection{New MASCARA candidate variable stars} \label{ss:new_res}

Finally the 45\,749 stars in the MASCARA sample with sufficient data that are not listed in the VSX catalogue were analysed. Periodic variations were detected in the light curves of 438 of these stars. Checks against the VSX catalogue for background variables (Step 4) revealed 156 false positives, leaving 282 as new MASCARA candidate variable stars listed in Table~\ref{t:n}. As with the known variable stars discussed in Sect.~\ref{ss:param}, an estimate of the type of variability of these stars (eclipsing binary, pulsating or other) was made based on their MASCARA light curves. 44 were visually identified as possible eclipsing binary systems.

Light curves and periodograms for seven example new candidate variables are included in Appendix~\ref{a:n}, and can be found for all candidates at \url{https://home.strw.leidenuniv.nl/~burggraaff/MASCARA_variables/}. The reader should note that these stars still need to be vetted with further observations.

An interesting example of a new candidate variable star is ASCC 201832 (\object{TYC 3926-224-1}). This star is not known in the extended literature to have a variable or binary nature. Though it is a relatively bright star ($V = 7.42$ mag), it was not included in the Hipparcos catalogue \citep{HIP_vars}. It has been included, but not flagged as a variable star, in the second Gaia data release \citep{GaiaDR2, GaiaDR2vars}.

\begin{figure}
	\centering
	\includegraphics[width=\columnwidth]{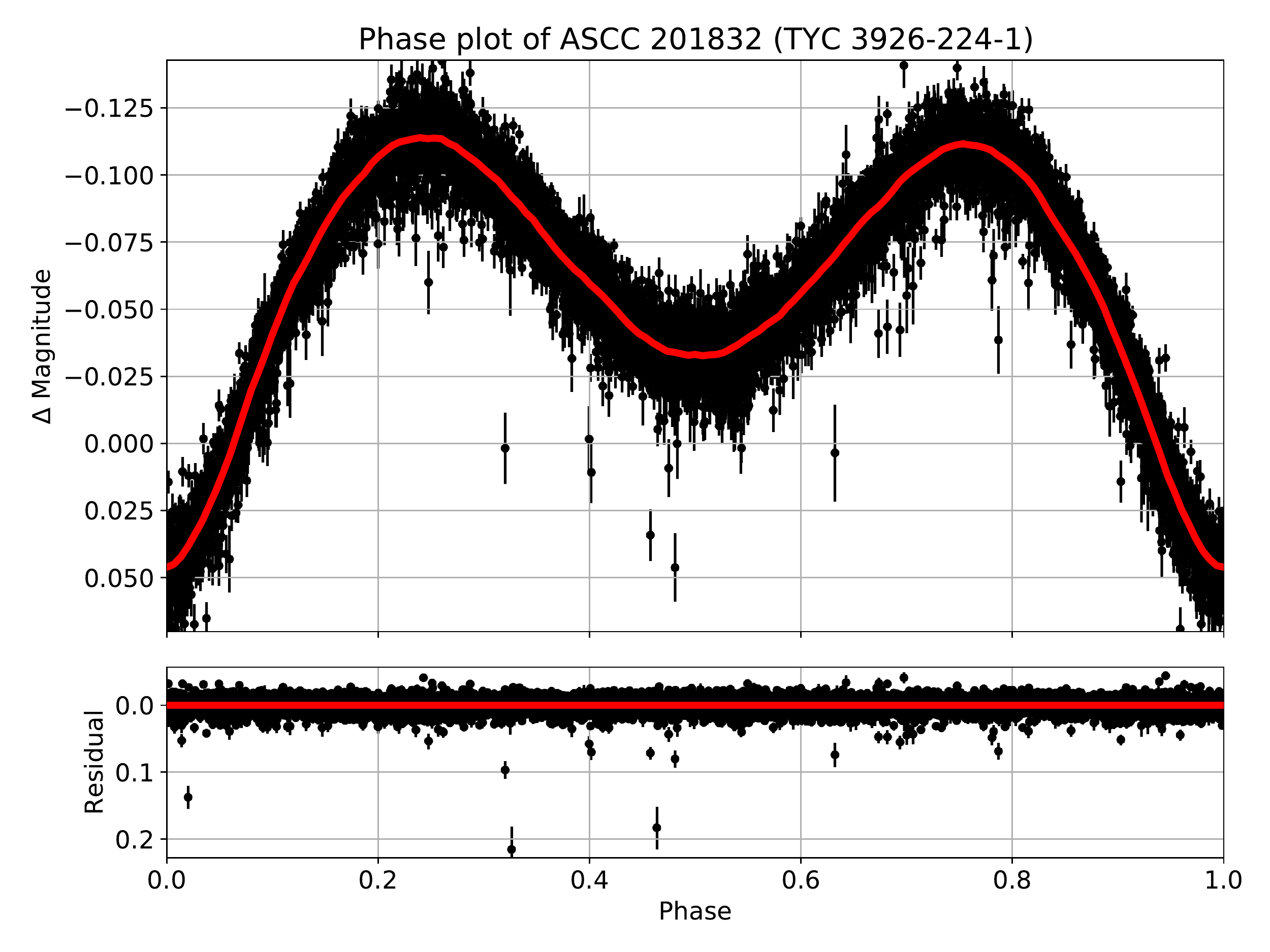}
	\caption{MASCARA light curve of ASCC 201832 (\object{TYC 3926-224-1}), a new variable star candidate with a period of $0.61747(5)$ days. The red line is the running average over an 0.025 phase interval. The light curve clearly shows a primary and secondary eclipse.}
	\label{f:201832}
\end{figure}

The MASCARA light curve of TYC 3926-224-1, given in Fig.~\ref{f:201832}, shows a clear variability with a period $P = 0.61747(5)$ days. The light curve is similar to that of $\beta$ Lyr type variables, with a primary and secondary eclipse, and a continuous change in brightness over the whole period. The depth of the primary eclipse is $160$ mmag, while the depth of the secondary eclipse is $81$ mmag.

No stars significantly brighter than TYC 3926-224-1 were found within a degree from it. There are only two stars with $V < 8$ mag within a radius of 40 arcmin, with separations of 8.6 and 9.0 arcmin. Neither of these stars, HD 173700 and HD 173605 respectively, is known or suspected to be variable. Since both are fainter than TYC 3926-224-1 and the separations are sufficiently large, it is unlikely that the variability seen in TYC 3926-224-1 is due to blending with either of these stars.

There is one suspected variable star within 10 arcmin of TYC 3926-224-1, namely \object{NSV 24573} (\object{HD 173603}). This star was flagged as variable in the Hipparcos catalogue but has not been flagged by Gaia \citep{HIP_vars,GaiaDR2}. Variability in this star was also detected with MASCARA, with a period of $3.63(2)$ days and an amplitude of $37$ mmag. Thus it is unlikely that blending with HD 173603 has caused the variability observed in TYC 3926-224-1.

Since no likely blending candidates were found, it can be concluded that TYC 3926-224-1 itself is likely a new variable star. Given the shape of its light curve, it is likely an eclipsing binary of the $\beta$ Lyr type. However, we stress that follow-up observations are necessary to confirm its being variable.

\subsection{Detailed variability studies}

We also investigated to what extent particularly second order effects in the light curves of variable stars can be studied using MASCARA data. For this we focus on the O'Connell effect in eclipsing binaries.

The O'Connell effect is an asymmetry in the brightness of the two maxima in the light curve of an eclipsing binary system, of which the physical cause is not yet well understood \citep{oconnell, oconnell_revis}. It occurs in eclipsing binaries of the $\beta$ Lyr and W UMa subtypes, with the maximum after the primary eclipse being brighter than that before the primary eclipse.

\begin{figure}
	\centering
	\includegraphics[width=\columnwidth]{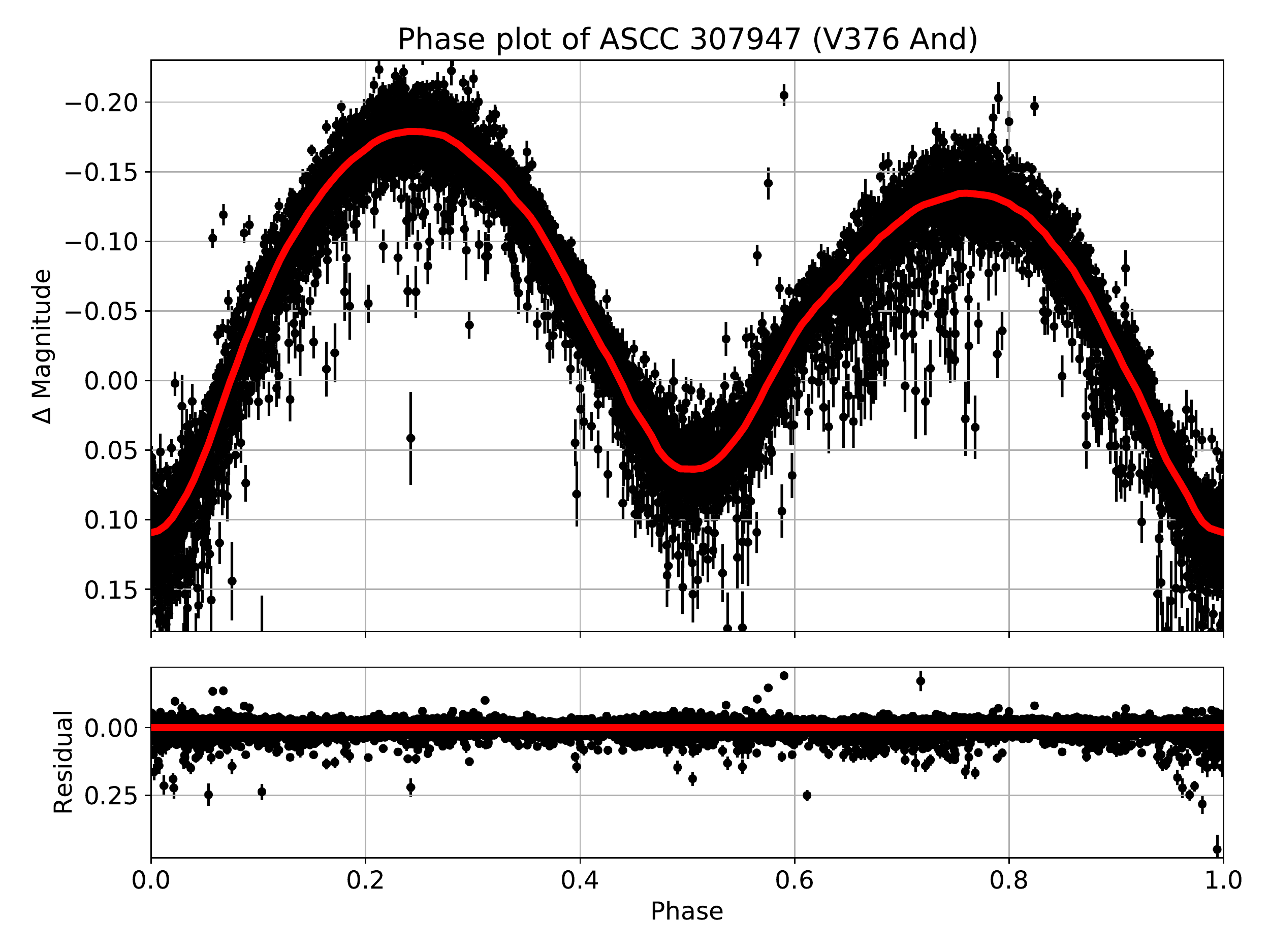}
	\caption{Light curve of ASCC 307947 (\object{V376 And}), a known $\beta$ Lyr type eclipsing binary with a period of 0.79867(6) days. The O'Connell effect is visible as a difference in magnitude between the two maxima. The red line is the running average over an 0.025 phase interval.}
	\label{f:ocon}
\end{figure}

An example MASCARA light curve of a star showing this effect, the $\beta$ Lyr variable \object{V376 And}, is given in Fig.~\ref{f:ocon}, with the first maximum approximately 0.05 mag brighter than the second. Table~\ref{t:ocon} contains a non-exhaustive list of known eclipsing binaries in the MASCARA sample exhibiting a significant O'Connell effect. The effect is not detected in any of the newly found variables.

\begin{table*}
	\caption{Eclipsing binaries in the MASCARA sample that exhibit the O'Connell effect. Periods include a 3$\sigma$ confidence interval. The amplitude is that of the full oscillation in the MASCARA band. $\Delta m$ is the difference in magnitude between the primary (after primary minimum) and secondary (before primary minimum) maxima in the binned light curve. No previous detections of the O'Connell effect in \object{HD 219561} and \object{V1392 Ori} were found.}
	\label{t:ocon}
	\centering
	\begin{tabular}{rllllllrl}
		\hline \hline
		ASCC        & Identifier      & V    & RA                         & Dec                            & Period   & Amplitude      & $\Delta m$ & Previous \\
		            &                 &      & (J2000)                    & (J2000)                        & (days)   & (mag)     & (mmag)     & detection \\
		\hline
		307947    & \object{V376 And}  & 7.77 & 02$^h$35\arcmin11.6\arcsec & $+$49\degree51\arcmin37\arcsec & 0.79867(6)  & 0.29 & 44    & (1) \\
		449928    & \object{HD 219561} \tablefootmark{a} & 8.40 & 23$^h$16\arcmin21.2\arcsec & $+$41\degree33\arcmin43\arcsec & 0.56660(5) & 0.28 & 21 & -- \\
		513514    & \object{V556 Lyr}  & 8.14 & 19$^h$25\arcmin08.3\arcsec & $+$35\degree59\arcmin58\arcsec & 1.4901(3)   & 0.13 & 13    & (2) \\
		521078    & \object{V448 Cyg}  & 8.14 & 20$^h$06\arcmin09.9\arcsec & $+$35\degree23\arcmin10\arcsec & 6.519(2)    & 0.37 & 18    & (3) \\
		558695    & \object{V600 Per}  & 7.86 & 03$^h$19\arcmin01.4\arcsec & $+$32\degree41\arcmin16\arcsec & 1.4697(1)   & 0.39 & 26    & (4) \\
		725559    & \object{ER Vul}    & 7.35 & 21$^h$02\arcmin25.9\arcsec & $+$27\degree48\arcmin26\arcsec & 0.69810(5)  & 0.13 & 13    & (5) \\
		1017908 & \object{V1392 Ori} & 7.76 & 06$^h$16\arcmin17.9\arcsec & $+$09\degree01\arcmin40\arcsec & 1.3881(1)   & 0.20 & 24    & -- \\
		\hline
	\end{tabular}
	\tablefoot{
		\tablefoottext{a}{Identified as \object{NSVS 6156390} in the ASCC-VSX cross-match; identified as \object{TYC 3225-1270-1} in the ASCC catalogue.}
	}
	\tablebib{
		(1) \citet{V376And}; (2) \citet{V556Lyr}; (3) \citet{V448Cyg}; (4) \citet{V600Per}; (5) \citet{ERVul}.
	}
\end{table*}

\section{Discussion and conclusions} \label{s:dis}

To our knowledge, MASCARA is the first instrument to accurately monitor the near-entire sky, recording the flux of all bright stars ($V < 8.4$ mag) down to airmass two to three every 6.4 seconds. Typical precisions of 1.5\% per five minutes are reached at the faint magnitude end. With the analysis presented here we show that MASCARA data are very well suited to study known variable stars and can serve as a powerful means to find new variables among the brightest stars in the sky.

Using a generalised Lomb-Scargle analysis we show that 93.5\% of all known variables with periods between 0.1 and 10 days and amplitudes $>$ 2\% are recovered using the first year of MASCARA data. However, great care has to be taken to remove systematic effects in the data, in particular with periods of one sidereal day and aliases thereof. Hence, identifying and studying stars that exhibit variability with a period at or near $1.0$ day will be very challenging with MASCARA data alone.

The recovery fraction of known variable stars drops significantly below periods of 0.1 day ($158 / 231 = 68\%$). We note that short period variables often show multi-periodicity and irregular light curves, which are therefore more challenging to detect using a Lomb-Scargle analysis. For the MASCARA sample, this is mostly relevant for $\delta$ Scuti stars \citep{delsct1, delsct2}, of which there are 360 in the ASCC-VSX cross matched catalogue. At long periods ($>10$ days), two main causes for the relatively low recovery rate ($425 / 623 = 68\%$) can be identified. Firstly, since the MASCARA data set spans 424 days, stars with long periods simply have fewer cycles in the data. This reduces the robustness against missing or bad data and instrumental effects, such as that caused by the moon. Secondly, many long-period variables also show multi-periodicity and irregularities in the shapes, amplitudes and periods of their light curves \citep{longperiods1, longperiods2, longperiods3, longperiods4}, which both make it more difficult to determine the main period over only a few cycles, and can cause the current period of the star to be inherently different from that measured previously. When searching for new unknown variable stars in the MASCARA data, as discussed below, the dependence of the recovery rate on amplitude and period has implications on the reliability of the parameters found.

From the whole test sample of 2776 known variables, 401 (14.4\%) are not recovered. Of these, 98 are classified in the VSX as irregular or semi-regular; the non-recovery of these stars can be explained intuitively by irregularities in their behaviour. A further 47 stars belong to classes of variable stars known to show multi-periodicity, which also easily explains their non-recovery. This leaves 256 stars of which the non-recovery cannot be easily explained by their class. For 69 of these, no variability larger than the typical scatter in the binned data points (0.03 magnitude) is detected. Additionally some cases of non-recovery can be explained by outliers in the data, aliasing, or incomplete filtering of either of the two systematics described in Sect.~\ref{ss:detrend}. Finally there are some examples of which the light curve appears to be well-described by the found period; these may be cases of previously unknown multi-periodicity, periods that have changed over time, or simply errors in the VSX catalogue. A more thorough analysis of the unrecovered known variables would be needed to assign individual stars to each of these classes.

The MASCARA data can also effectively be used to further characteri{s}e known variable stars. For example, for 210 stars the MASCARA data determines for the first time a period. By detecting the O’Connell effect in several eclipsing binaries, we show that MASCARA data are of sufficient quality to study second order variability effects in the brightest stars in the sky. 

As presented in Sects.~\ref{ss:param} and~\ref{ss:new_res}, we have determined new parameters for 210 known variables from the VSX catalogue, shown in Table~\ref{t:p} in Appendix~\ref{a:p}, and identified 282 new candidate variable stars, which are presented in Table~\ref{t:n} in Appendix~\ref{a:n}. We note that this only means that these stars either are not present in the VSX catalogue, or are present but lack parameters, and that some may already have been recorded in the extended literature. Although the new candidates have been vetted for possible known background variable stars in the VSX database that could cause the observed variability, they need to be observed with camera systems with significantly smaller pixel scales to exclude the contribution from possible faint unknown variable stars within the 2.5\arcmin{} aperture used for the MASCARA photometry and beyond.

The recovery rate discussed in Sect.~\ref{ss:acc} and its dependence on the period of variability are important to take into account. For stars with periods $< 0.1$ and $\gtrsim 10$ days, the recovery rate of known variables is as low as 68\%, casting doubt on the accuracy of the parameters determined with MASCARA for such stars. However, there are also mismatches between VSX and ASCC that are not only due to issues with the data or analysis. For example, there are stars that have multiple periods or irregular variability. Follow-up, either with more MASCARA data or with different instruments, can clear up the accuracy of the new parameters.

We note that our methods are sensitive to periodic variable stars, not to non-periodic ones. An alternative analysis, for instance a comparison with nearby stars, may be suited to find and characterise such stars. Additionally, the analysis is only performed on known stars in the ASCC master catalogue. This means that such objects as novae and flare stars, which were faint when the catalogue was created but can reach MASCARA magnitudes at later times, will not be detected with our analysis.

One interesting question that remains is why some of the new candidates have not previously been seen to be variable. For instance, ASCC 201832 (\object{TYC 3926-224-1}) has a V-magnitude of 7.42, a period of 0.61747(5) days and an amplitude of 160 mmag. Its light curve, given in Fig.~\ref{f:201832}, shows a very clear and regular variability, and one could reasonably expect this variable star to have been noticed earlier. A possible explanation for this lack of previous detection, as discussed in Sect.~\ref{s:intro}, is the fact that many previous surveys have focused on fainter stars than MASCARA and thus these stars may simply have `slipped through'. Additionally, many of the new variables have very short ($< 0.1$ days) or very long ($>10$ days) periods. Previous studies may not have had the cadence or duration necessary to detect such variability.

\begin{acknowledgements}
This project has received funding from the European Research Council (ERC) under the European Union's Horizon 2020 research and innovation programme (grant agreement nr. 694513).
\end{acknowledgements}

\bibliographystyle{aa}
\bibliography{paper_references}

\begin{appendix}
\section{New parameters of known and suspected variables} \label{a:p}

The GLS periodograms (top panel) and phase folded light curves (bottom panel) of seven example known variable stars with new parameters from MASCARA are provided here. These figures can be found for all 210 such stars at \url{https://home.strw.leidenuniv.nl/~burggraaff/MASCARA_variables/}. \\

\textbf{\emph{Notes for Table~\ref{t:p}.}} \\
$^a$ Variability types in the `MASCARA' column were visually estimated based on the shape, period and amplitude of the light curve. These are only included if they disagree with or complement the VSX catalogue. `E' indicates an eclipsing binary, `P' a pulsating variable. \\
$^b$ Errors correspond to a $3 \sigma$ confidence interval. For entries marked with a colon (:), no such confidence interval could be determined -- these are listed with two significant digits. \\
$^c$ Amplitude in the MASCARA band. \\
$^d$ Epoch of a minimum in brightness.

\begin{figure}
	\centering
	\includegraphics[width=\columnwidth]{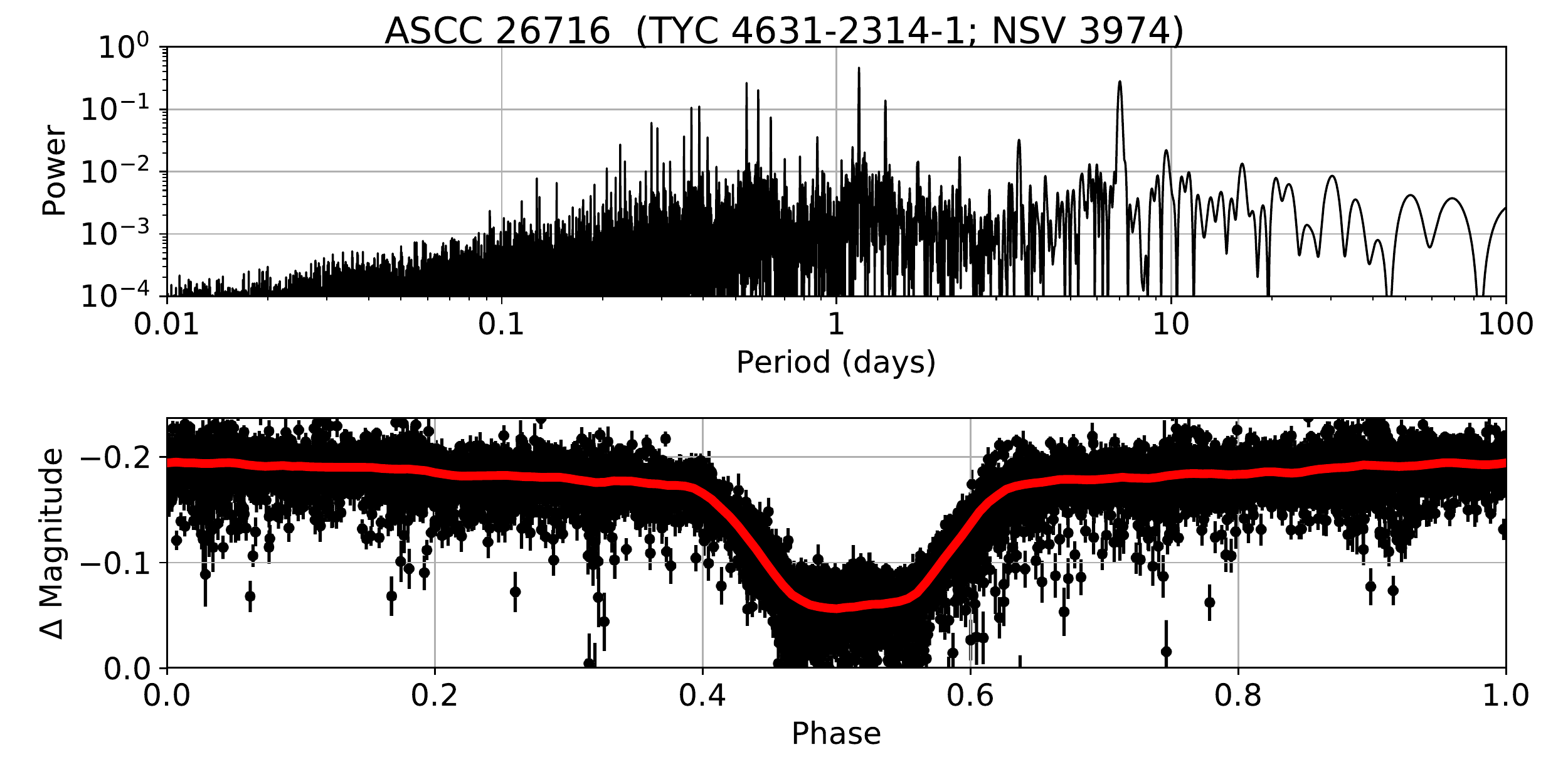}
	\caption{ASCC 26716 (\object{TYC 4631-2314-1}; \object{NSV 3974}); $p =$ 1.1664(2) d.}
\end{figure}

\begin{figure}
	\centering
	\includegraphics[width=\columnwidth]{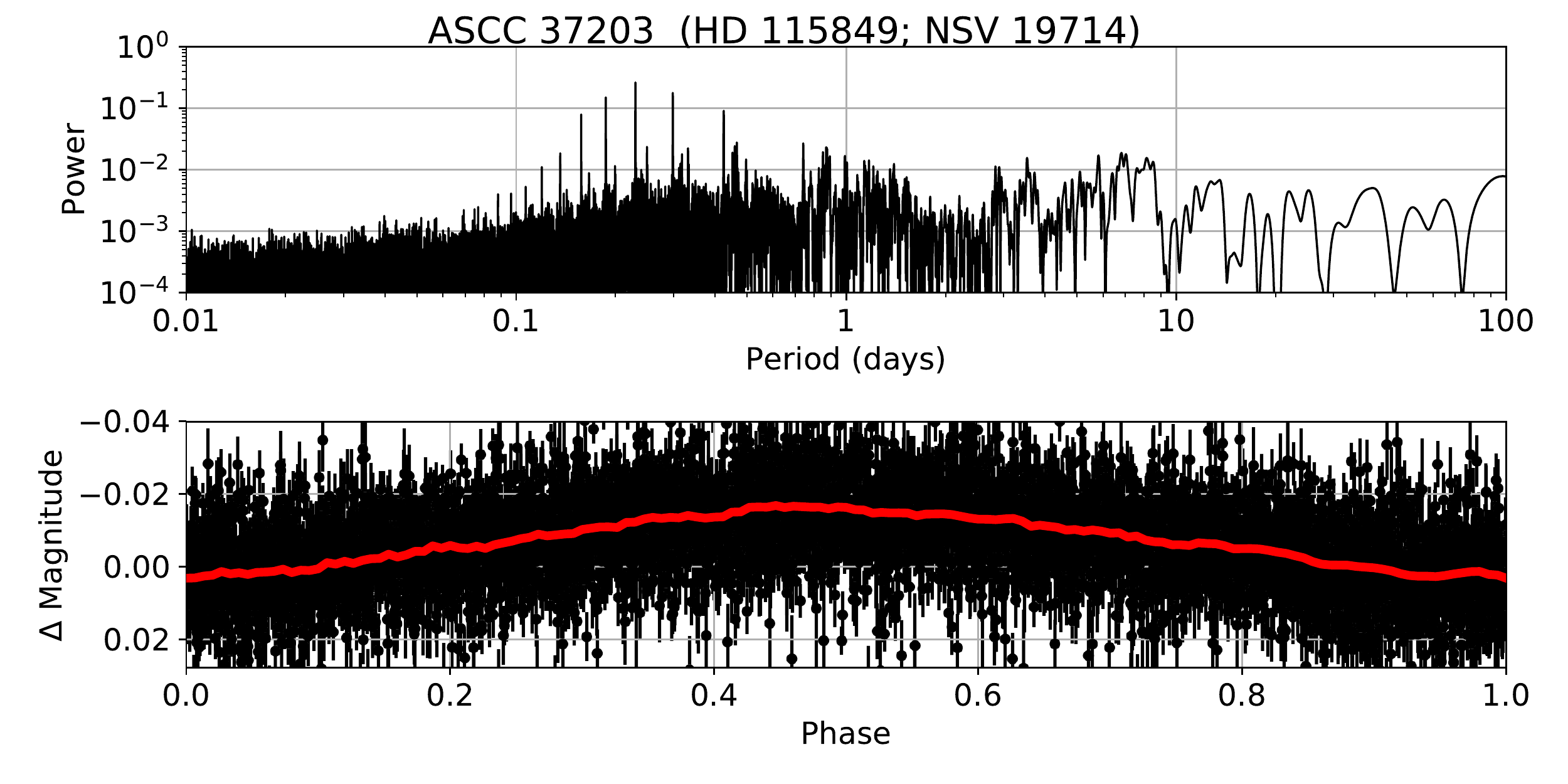}
	\caption{ASCC 37203 (\object{HD 115849}; \object{NSV 19714}); $p =$ 0.2297(2) d.}
\end{figure}

\begin{figure}
	\centering
	\includegraphics[width=\columnwidth]{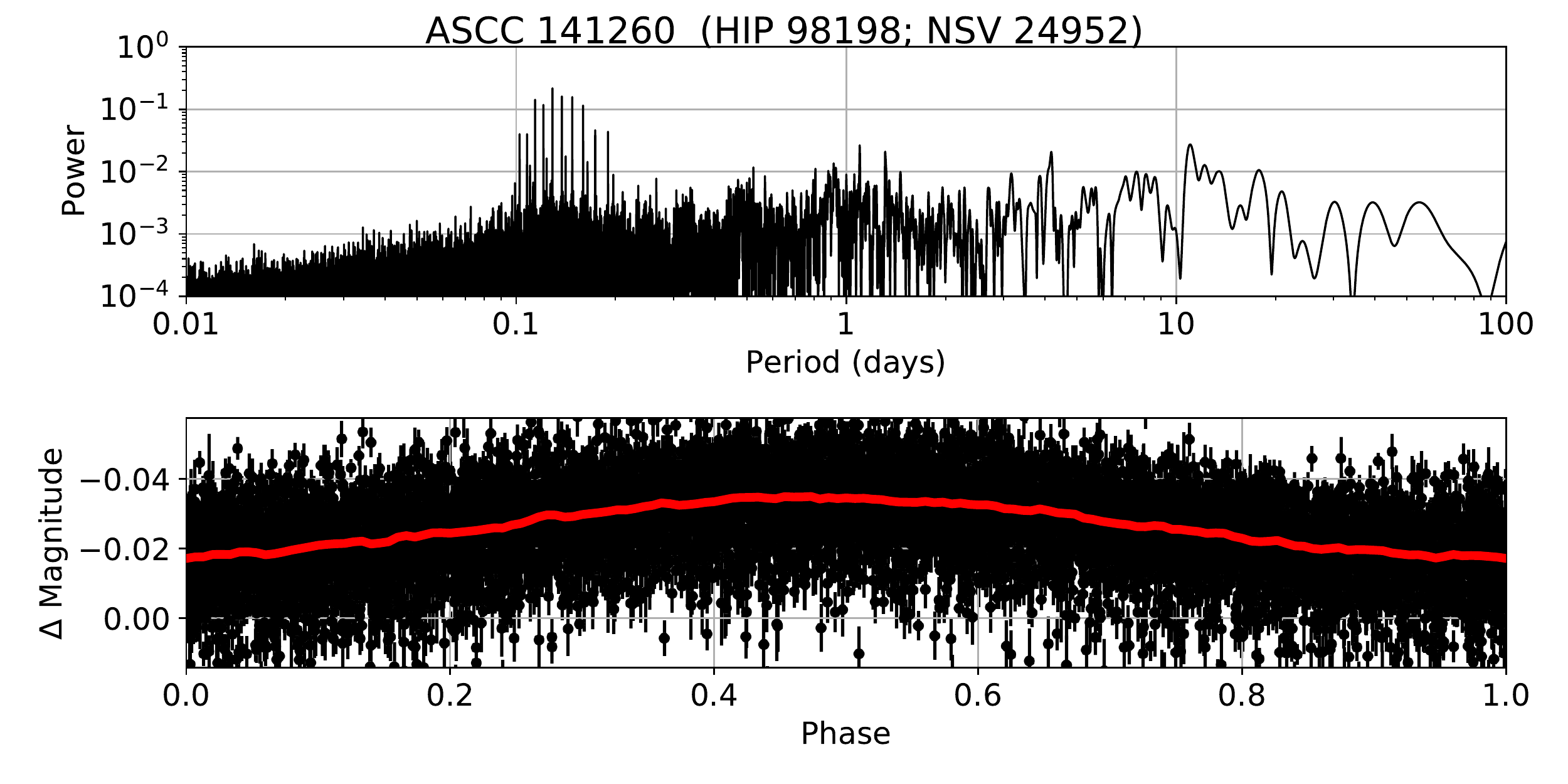}
	\caption{ASCC 141260 (\object{HIP 98198}; \object{NSV 24952}); $p =$ 0.12879(4) d.}
\end{figure}

\begin{figure}
	\centering
	\includegraphics[width=\columnwidth]{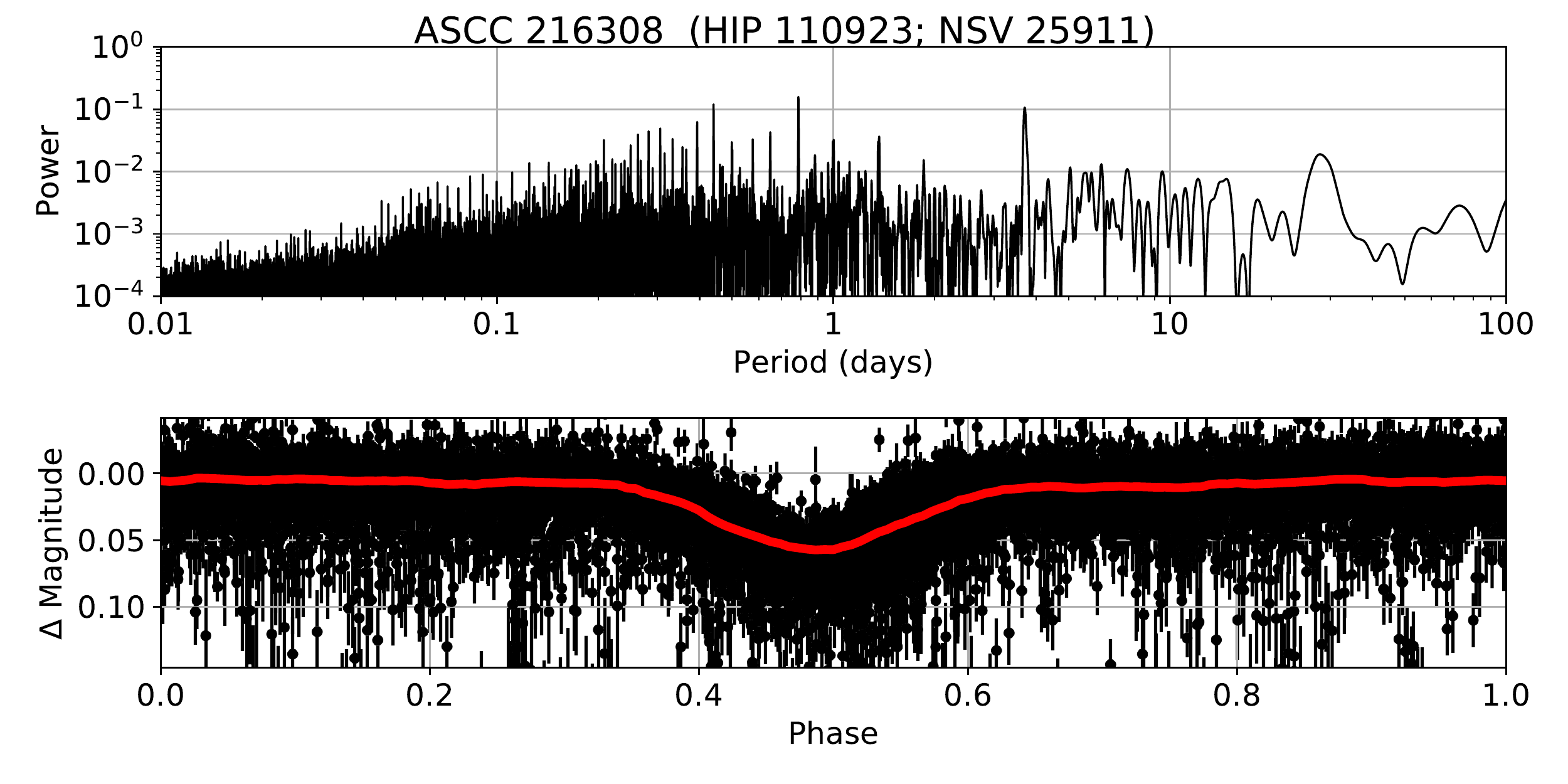}
	\caption{ASCC 216308 (\object{HIP 110923}; \object{NSV 25911}); $p =$ 0.7866(6) d.}
\end{figure}

\begin{figure}
	\centering
	\includegraphics[width=\columnwidth]{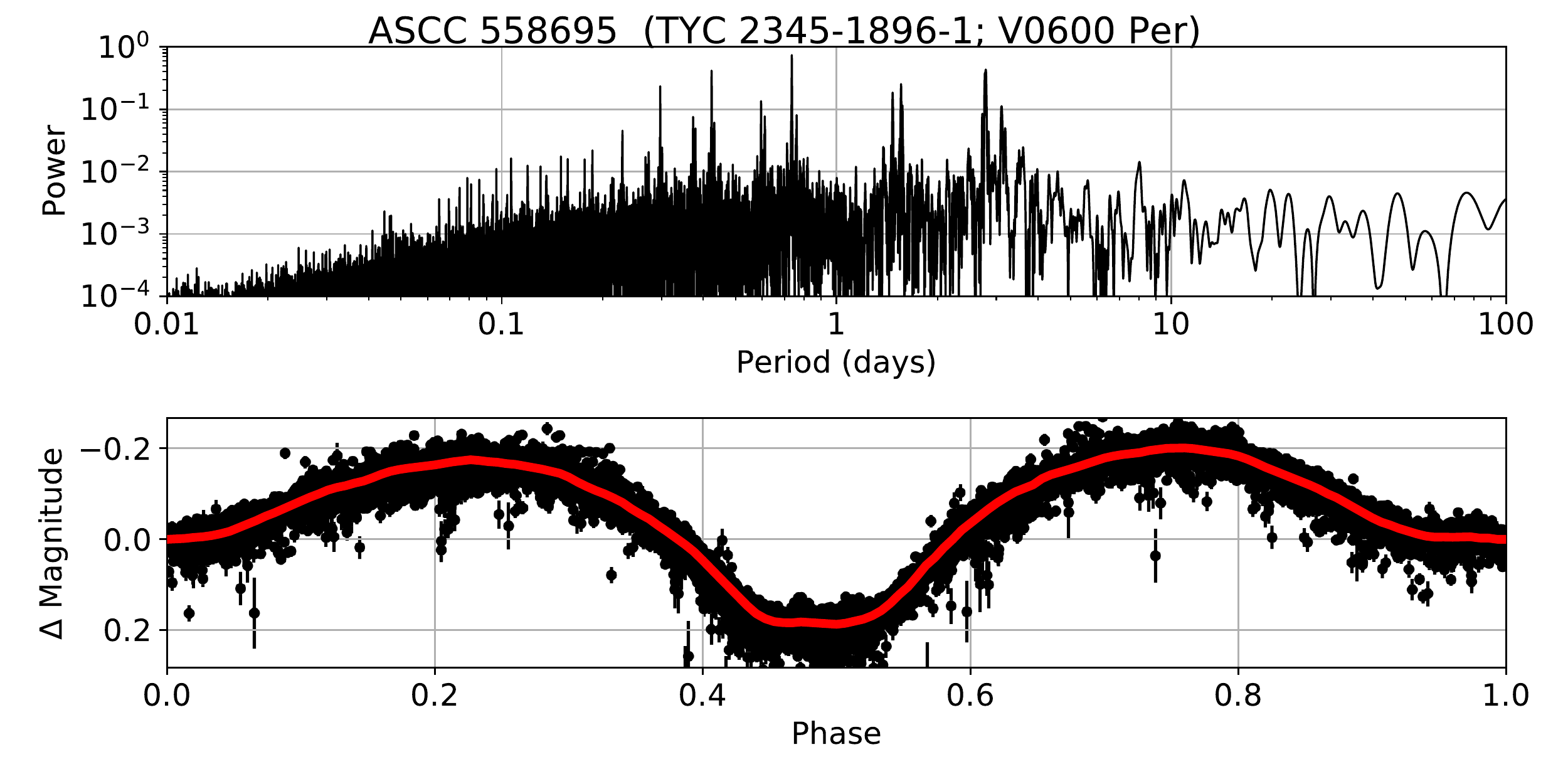}
	\caption{ASCC 558695 (\object{TYC 2345-1896-1}; \object{V0600 Per}); $p =$ 1.4697(1) d.}
\end{figure}

\begin{figure}
	\centering
	\includegraphics[width=\columnwidth]{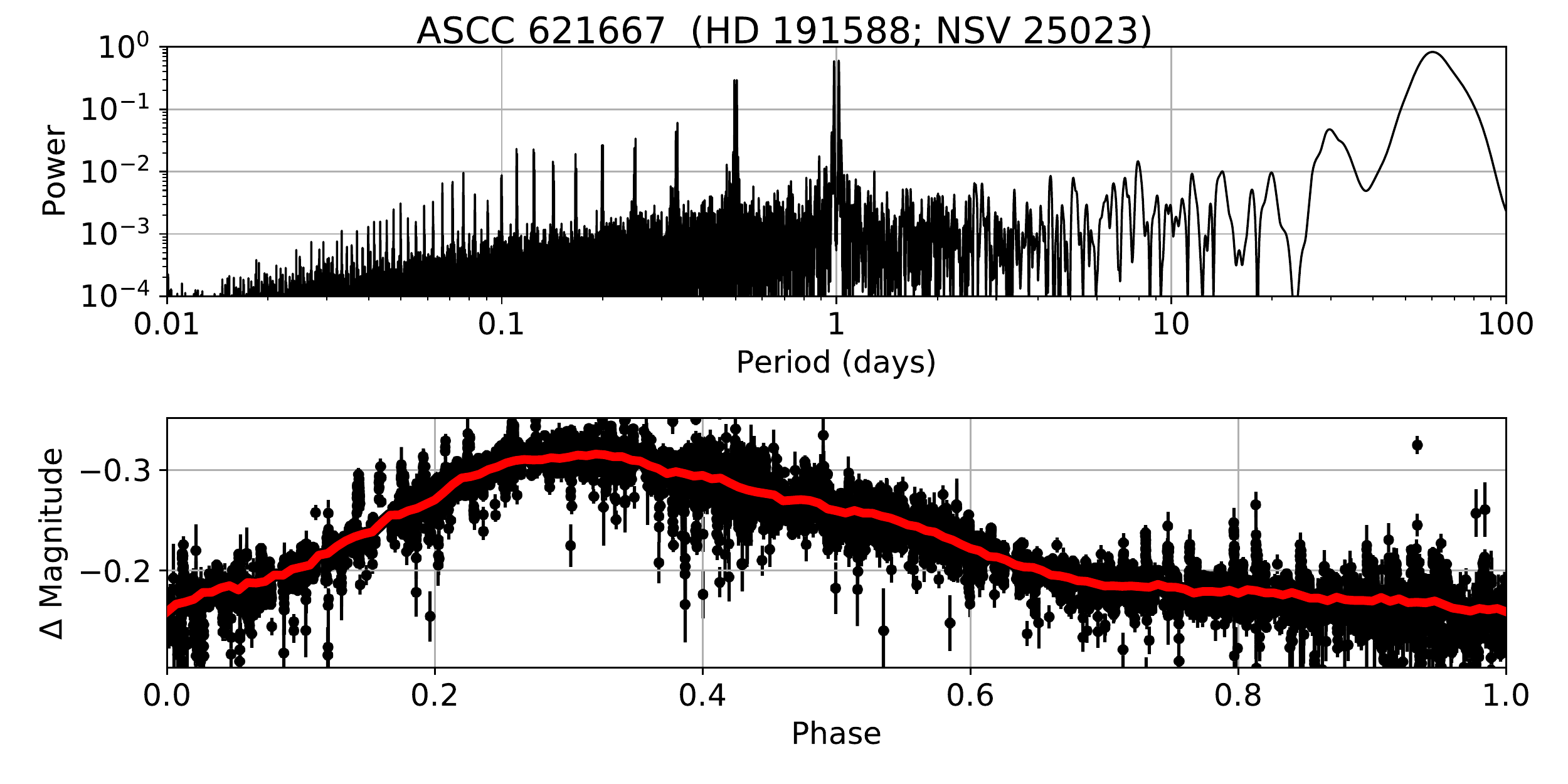}
	\caption{ASCC 621667 (\object{HD 191588}; \object{NSV 25023}); $p =$ 61(1) d.}
\end{figure}

\begin{figure}
	\centering
	\includegraphics[width=\columnwidth]{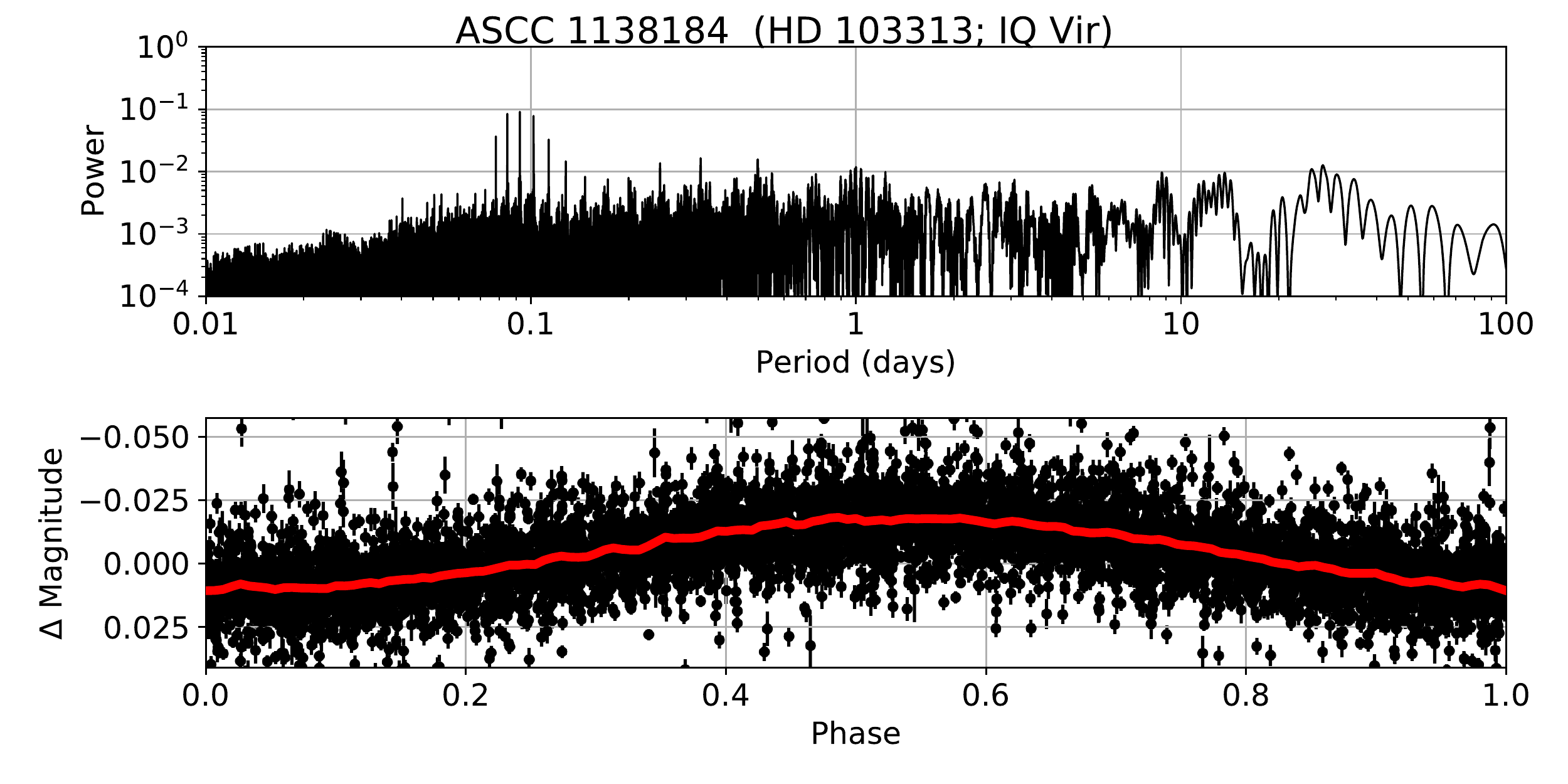}
	\caption{ASCC 1138184 (\object{HD 103313}; \object{IQ Vir}); $p =$ 0.092463(3) d.}
\end{figure}

\longtab[1]{
\begin{landscape}
\begin{longtable}{rlllllllrlrl}
\caption{\label{t:p}Known and suspected variable stars from the VSX catalogue without well-determined parameters (period and amplitude) detected with MASCARA. \\ The full version of this table is also available in electronic form at the CDS via anonymous ftp to cdsarc.u-strasbg.fr (130.79.128.5) or via \url{http://cdsweb.u-strasbg.fr/cgi-bin/qcat?J/A+A/}.} \\
\hline
\hline
ASCC & Identifier & VSX ID & \multicolumn{2}{c}{Var. Type\tablefootmark{a}} & V & RA (J2000) & Dec (J2000) & Nr. Obs. & Period\tablefootmark{b}  & Amp.\tablefootmark{c} & Epoch\tablefootmark{d} \\
 & & & VSX & MASCARA & & & & & (days) & (mmag) & (HJD $- 2\,450\,000$) \\
\hline
\endfirsthead
\caption{Continued.} \\
\hline
ASCC & Identifier & VSX ID & \multicolumn{2}{c}{Var. Type\tablefootmark{a}} & V & RA (J2000) & Dec (J2000) & Nr. Obs. & Period\tablefootmark{b}  & Amp.\tablefootmark{c} & Epoch\tablefootmark{d} \\
 & & & VSX & MASCARA & & & & & (days) & (mmag) & (HJD $- 2\,450\,000$) \\
\hline
\endhead
\hline
\endfoot
\hline
\endlastfoot
7\,027 & \object{HIP 9494} & \object{V0779 Cas} & EA: & -- & 6.69 & 02$^h$02\arcmin09.3\arcsec & $+$75\degree30\arcmin08\arcsec & 8857 & 6.3530(7) & 263 & 7203.5844 \\ 
7\,512 & \object{HIP 10309} & \object{NSV 15448} & DSCTC: & -- & 6.45 & 02$^h$12\arcmin49.9\arcsec & $+$79\degree41\arcmin30\arcsec & 9574 & 0.21: & 7 & 7227.72 \\ 
11\,321 & \object{HD 21179} & \object{V0805 Cas} & SRB & -- & 6.39 & 03$^h$30\arcmin19.4\arcsec & $+$71\degree51\arcmin50\arcsec & 7274 & 39.4(1) & 110 & 7240.7 \\ 
18\,914 & \object{HIP 26940} & \object{NSV 16615} & -- & P & 6.77 & 05$^h$43\arcmin00.7\arcsec & $+$73\degree59\arcmin09\arcsec & 5912 & 0.088: & 48 & 7201.718 \\ 
22\,011 & \object{HD 47505} & \object{EP Cam} & SRB & -- & 8.07 & 06$^h$48\arcmin37.5\arcsec & $+$76\degree59\arcmin24\arcsec & 6223 & 39.6(1) & 194 & 7234.7 \\ 
22\,755 & \object{HIP 34101} & \object{CM Cam} & FKCOM & -- & 6.99 & 07$^h$04\arcmin15.1\arcsec & $+$75\degree24\arcmin41\arcsec & 5541 & 15.87(6) & 65 & 7377.4 \\ 
24\,395 & \object{HIP 36978} & \object{NSV 17498} & -- & -- & 7.73 & 07$^h$36\arcmin02.2\arcsec & $+$76\degree50\arcmin00\arcsec & 6165 & 23.1(4) & 58 & 7306.7 \\ 
26\,716 & \object{TYC 4631-2314-1} & \object{NSV 3974} & -- & E & 8.31 & 08$^h$25\arcmin12.0\arcsec & $+$84\degree00\arcmin51\arcsec & 11785 & 1.1664(2) & 139 & 7280.4131 \\
\end{longtable}
\end{landscape}
}

\clearpage

\section{New variable star candidates} \label{a:n}
The GLS periodograms (top panel) and phase folded light curves (bottom panel) of seven example candidate new variable stars from MASCARA are provided here. These figures can be found for all 282 such stars at \url{https://home.strw.leidenuniv.nl/~burggraaff/MASCARA_variables/}. \\

\textbf{\emph{Notes for Table~\ref{t:n}.}} \\
$^a$ Variability types were visually estimated based on the shape, period and amplitude of the light curve. `E' indicates an eclipsing binary, `P' a pulsating variable. \\
$^b$ Errors correspond to a $3 \sigma$ confidence interval. For entries marked with a colon (:), no such confidence interval could be determined -- these are listed with two significant digits. \\
$^c$ Amplitude in the MASCARA band. \\
$^d$ Epoch of a minimum in brightness.

\begin{figure}
	\centering
	\includegraphics[width=\columnwidth]{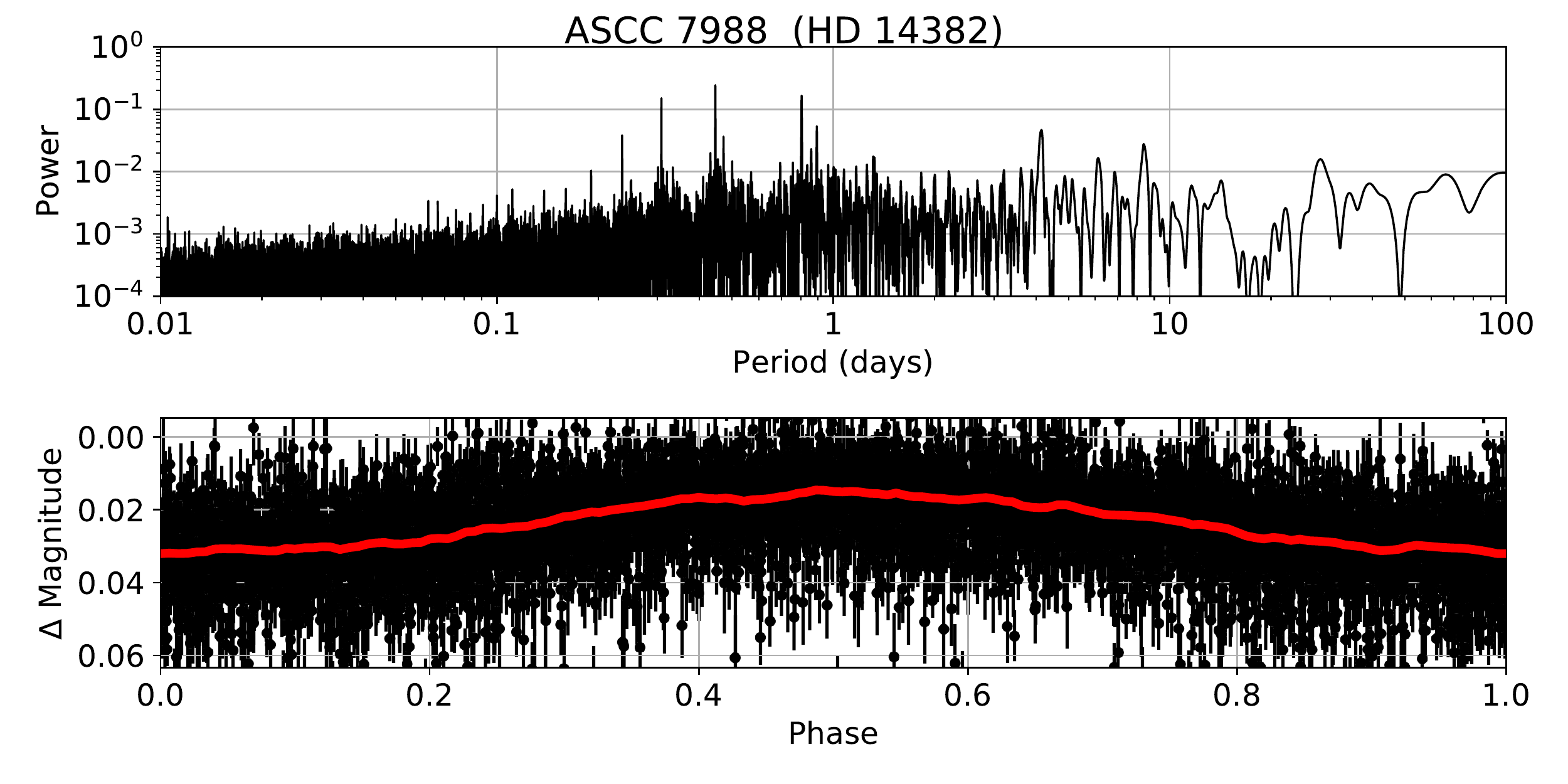}
	\caption{ASCC 7988 (\object{HD 14382}); $p =$ 0.4457(8) d.}
\end{figure}

\begin{figure}
	\centering
	\includegraphics[width=\columnwidth]{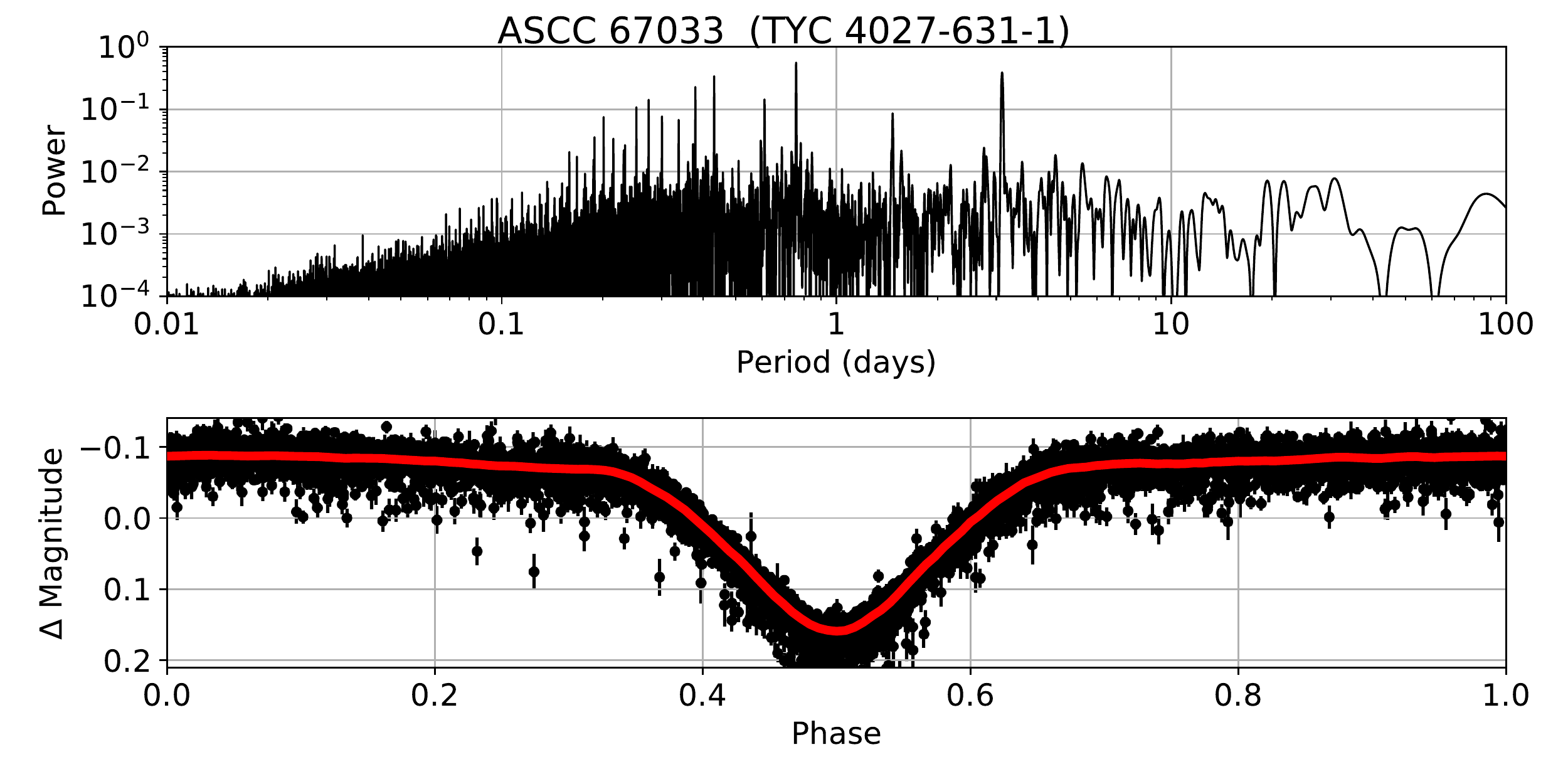}
	\caption{ASCC 67033 (\object{TYC 4027-631-1}); $p =$ 0.7569(1) d.}
\end{figure}

\begin{figure}
	\centering
	\includegraphics[width=\columnwidth]{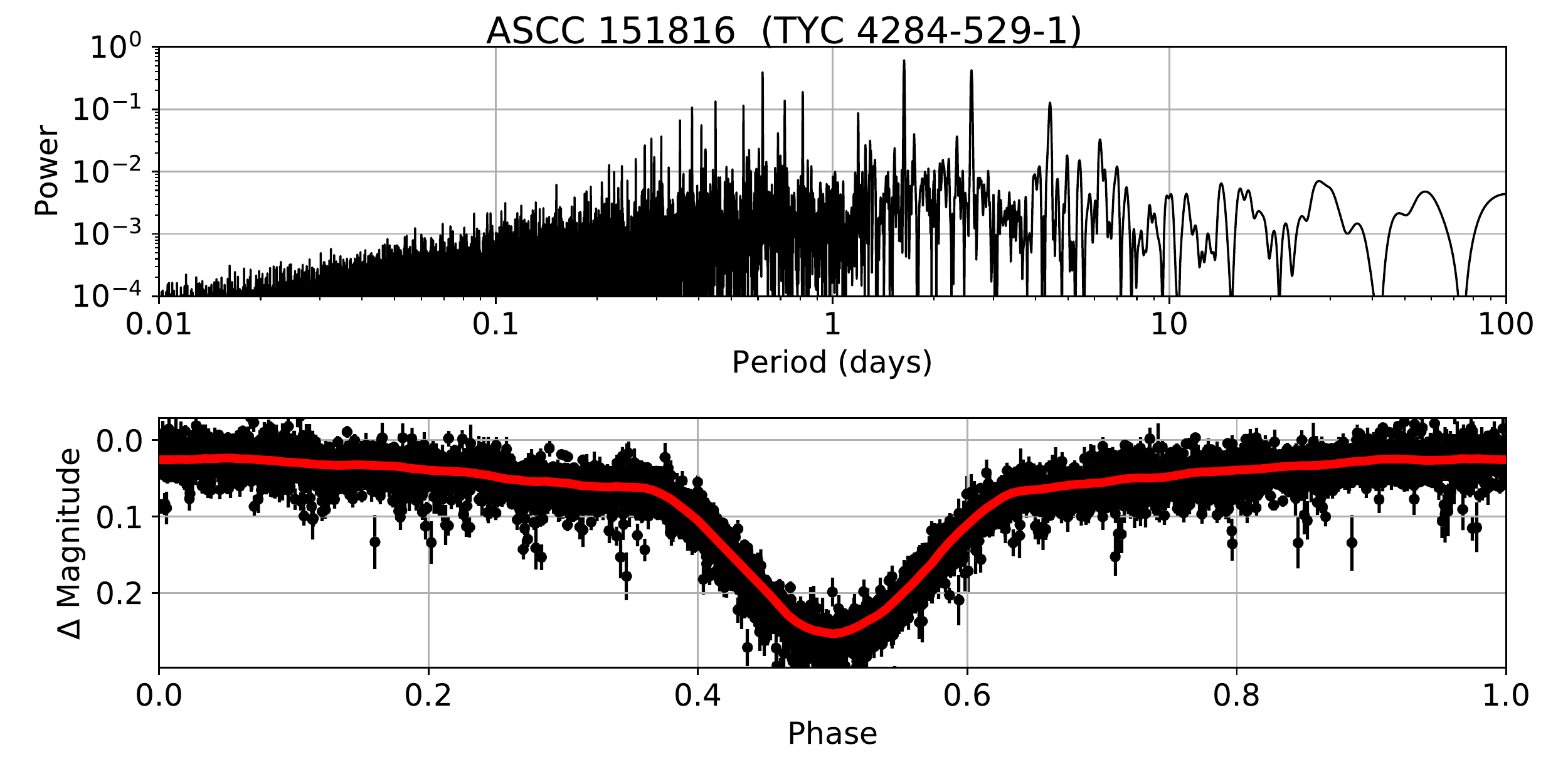}
	\caption{ASCC 151816 (\object{TYC 4284-529-1}); $p =$ 1.6295(5) d.}
\end{figure}

\begin{figure}
	\centering
	\includegraphics[width=\columnwidth]{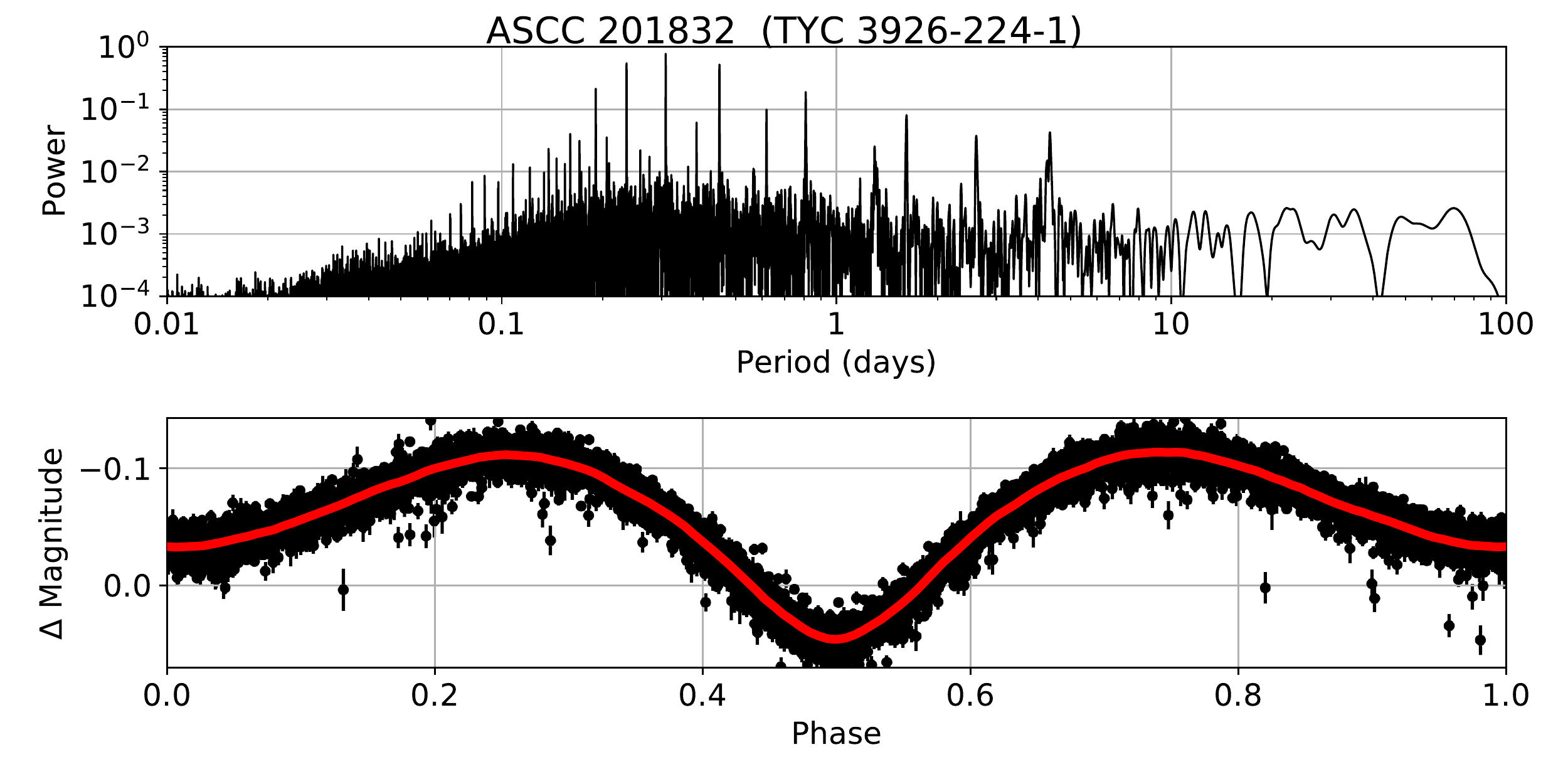}
	\caption{ASCC 201832 (\object{TYC 3926-224-1}); $p =$ 0.61747(6) d.}
\end{figure}

\begin{figure}
	\centering
	\includegraphics[width=\columnwidth]{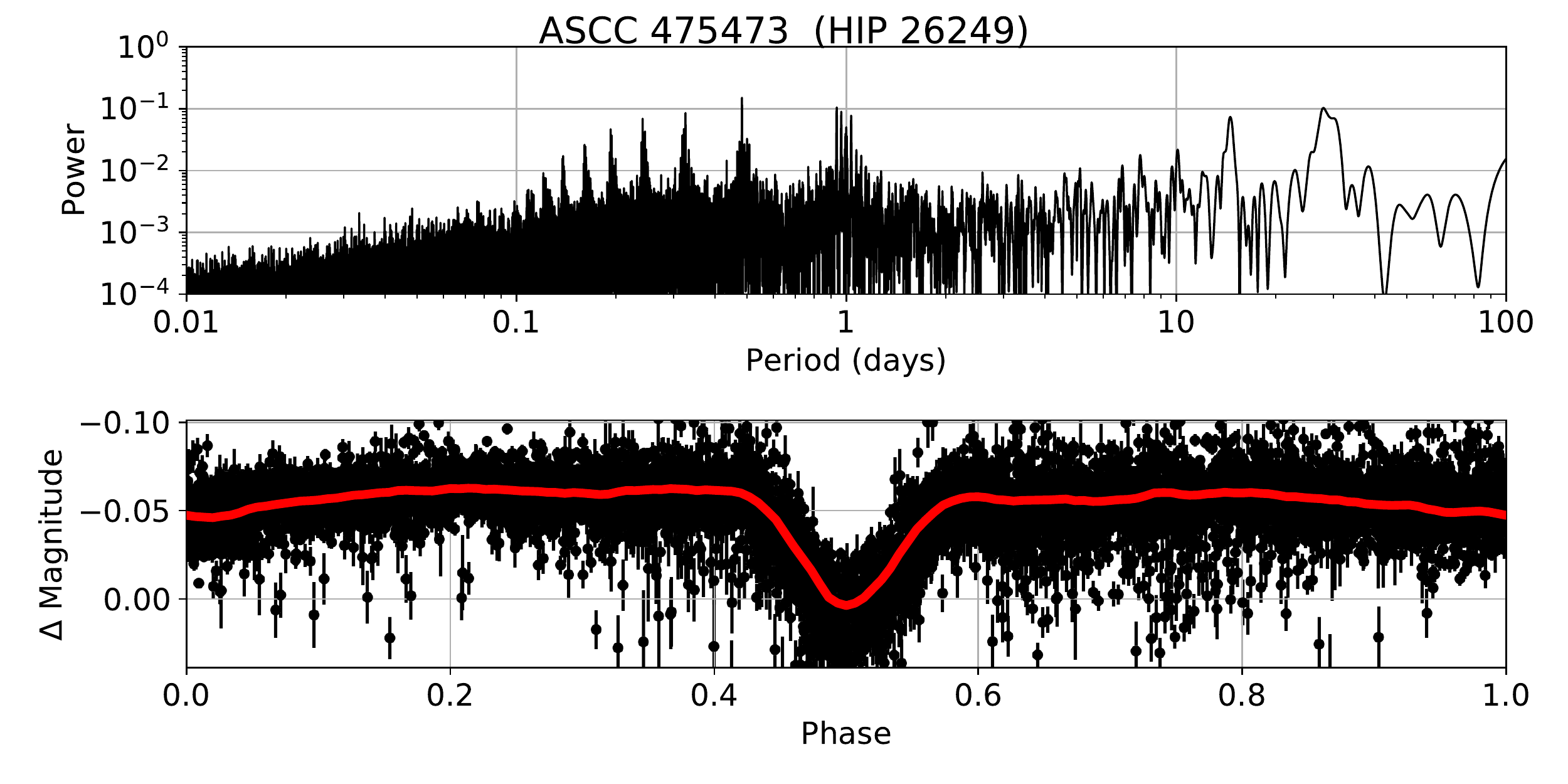}
	\caption{ASCC 475473 (\object{HIP 26249}); $p =$ 0.9656(2) d.}
\end{figure}

\begin{figure}
	\centering
	\includegraphics[width=\columnwidth]{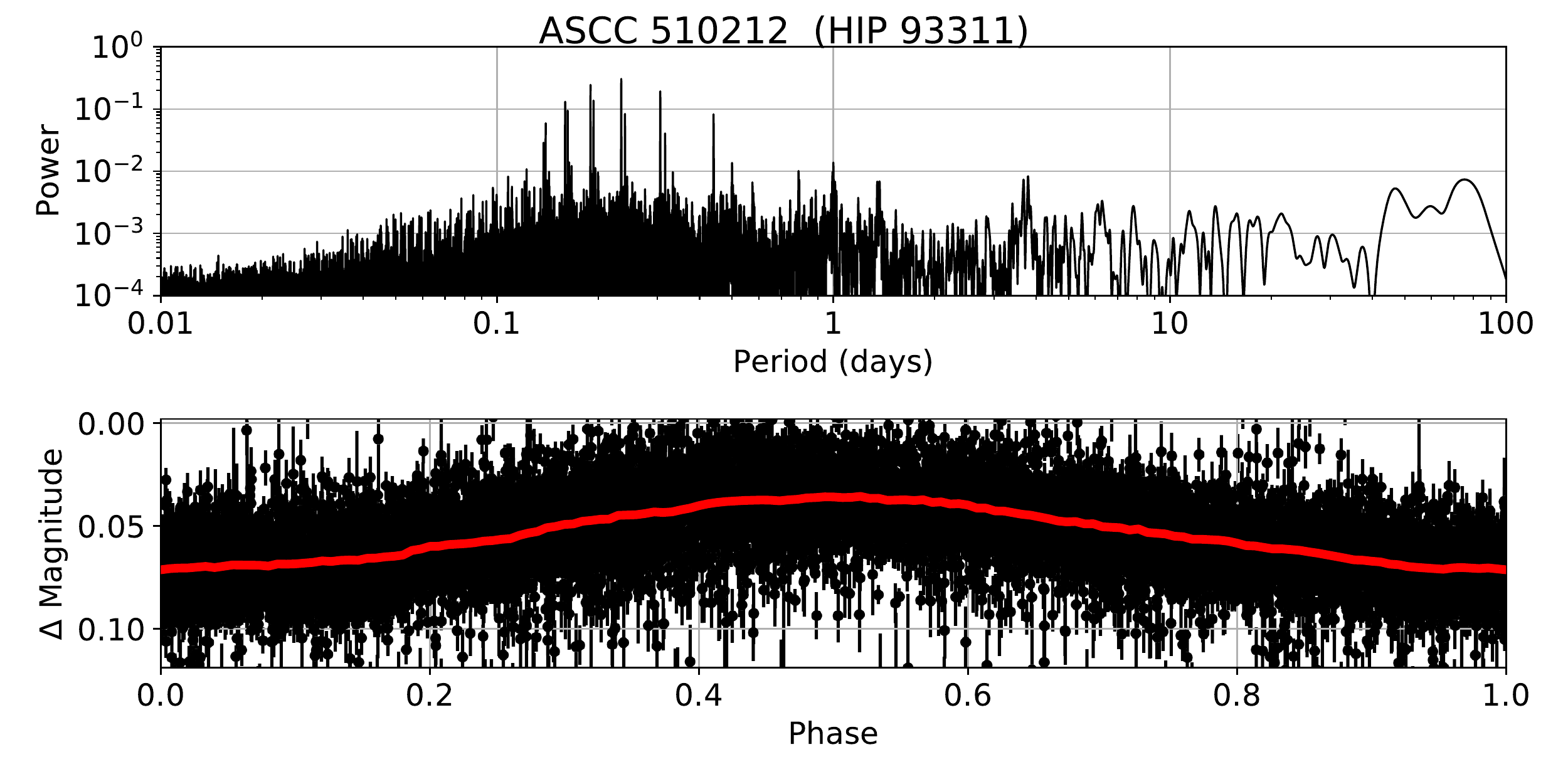}
	\caption{ASCC 510212 (\object{HIP 93311}); $p =$ 0.23392(6) d.}
\end{figure}

\begin{figure}
	\centering
	\includegraphics[width=\columnwidth]{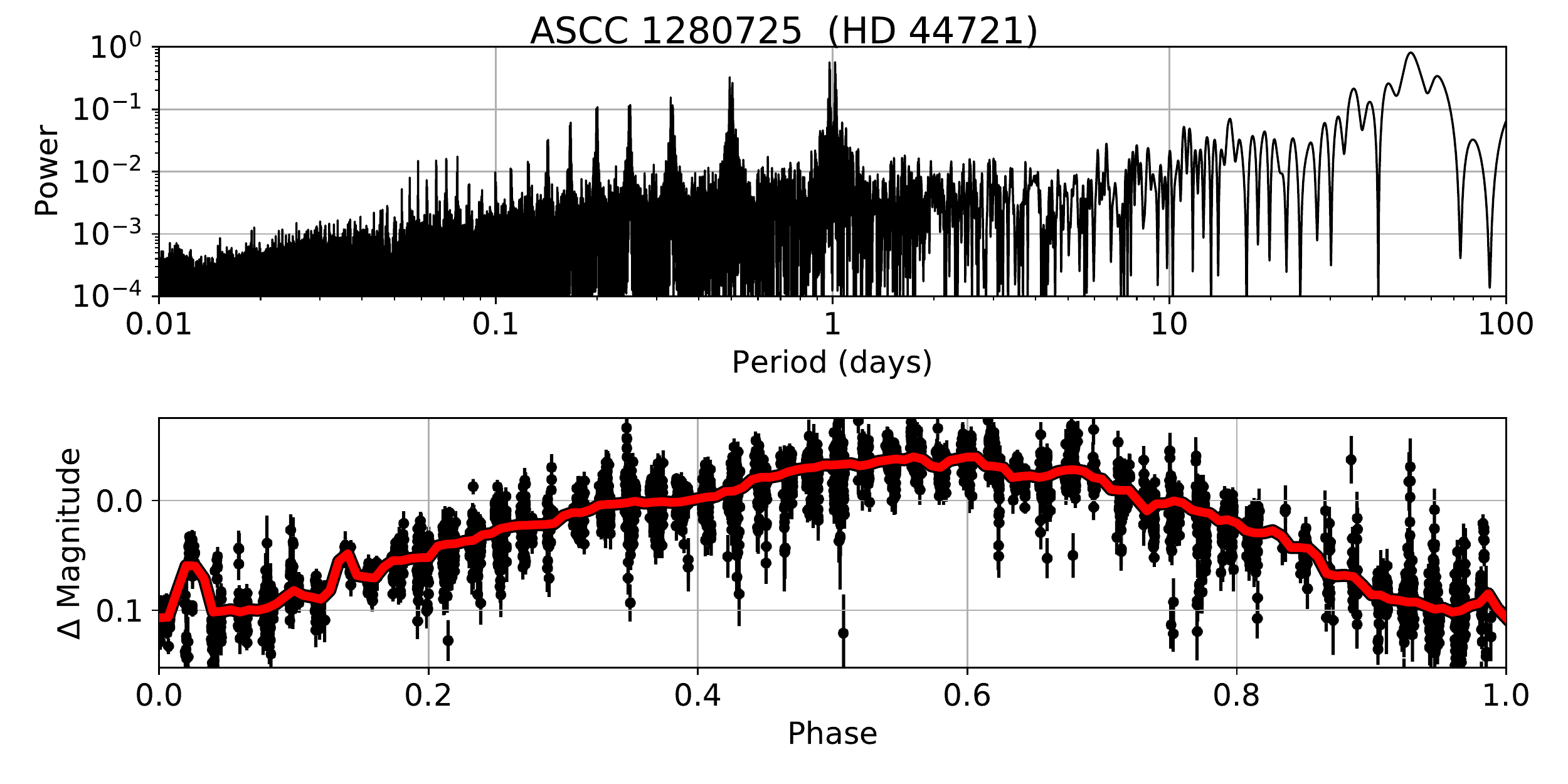}
	\caption{ASCC 1280725 (\object{HD 44721}); $p =$ 52.0(7) d.}
\end{figure}

\longtab[1]{
\begin{landscape}
\begin{longtable}{rlllllrlrl}
\caption{\label{t:n}New candidate variable stars detected with MASCARA. \\ The full version of this table is also available in electronic form at the CDS via anonymous ftp to cdsarc.u-strasbg.fr (130.79.128.5) or via \url{http://cdsweb.u-strasbg.fr/cgi-bin/qcat?J/A+A/}} \\
\hline
\hline
ASCC & Identifier & Var. Type\tablefootmark{a} & V & RA (J2000) & Dec (J2000) & Nr. Obs. & Period\tablefootmark{b}  & Amp.\tablefootmark{c} & Epoch\tablefootmark{d} \\
 & & & & & & & (days) & (mmag) & (HJD $- 2\,450\,000$) \\
\hline
\endfirsthead
\caption{Continued.} \\
\hline
ASCC & Identifier & Var. Type\tablefootmark{a} & V & RA (J2000) & Dec (J2000) & Nr. Obs. & Period\tablefootmark{b}  & Amp.\tablefootmark{c} & Epoch\tablefootmark{d} \\
 & & & & & & & (days) & (mmag) & (HJD $- 2\,450\,000$) \\
\hline
\endhead
\hline
\endfoot
\hline
\endlastfoot
3\,132 & \object{HIP 4090} & -- & 7.72 & 00$^h$52\arcmin29.0\arcsec & $+$79\degree50\arcmin26\arcsec & 10048 & 8.00(3) & 52 & 7306.77 \\ 
3\,347 & \object{TYC 4619-1223-1} & P & 8.14 & 00$^h$56\arcmin10.8\arcsec & $+$85\degree27\arcmin33\arcsec & 11238 & 0.094: & 18 & 7292.544 \\ 
7\,335 & \object{TYC 4503-2063-1} & P & 7.0 & 02$^h$08\arcmin56.1\arcsec & $+$79\degree20\arcmin45\arcsec & 9294 & 0.5414(7) & 16 & 7279.5832 \\ 
7\,988 & \object{HD 14382} & P & 7.82 & 02$^h$23\arcmin04.7\arcsec & $+$70\degree20\arcmin36\arcsec & 7800 & 0.4457(8) & 17 & 7279.6429 \\ 
11\,298 & \object{HD 20110} & -- & 8.36 & 03$^h$29\arcmin50.9\arcsec & $+$84\degree02\arcmin20\arcsec & 12276 & 33(1) & 69 & 7212 \\ 
20\,609 & \object{TYC 4525-994-1} & -- & 8.4 & 06$^h$17\arcmin55.6\arcsec & $+$76\degree40\arcmin29\arcsec & 5497 & 82: & 193 & 7269 \\ 
23\,934 & \object{HIP 36159} & E & 7.38 & 07$^h$26\arcmin57.6\arcsec & $+$75\degree36\arcmin44\arcsec & 5942 & 3.68(2) & 24 & 7349.78 \\ 
28\,165 & \object{HD 75544} & P & 7.32 & 08$^h$58\arcmin22.7\arcsec & $+$78\degree08\arcmin45\arcsec & 5446 & 0.11: & 13 & 7153.37 \\
\end{longtable}
\end{landscape}
}

\end{appendix}

\end{document}